\documentclass{article}
\usepackage{amsfonts}

\usepackage{graphicx}
\usepackage{amsmath}

\newtheorem{theorem}{Theorem}

\newtheorem{proposition}[theorem]{Proposition}

\newenvironment{proof}[1][Proof]{\textbf{#1.} }{\ \rule{0.5em}{0.5em}}
\makeatletter

\begin{document}

\title{{\small Available also at www.damianobrigo.it }\\ {\large Updated version to appear in Mathematical Finance} \\ --- \\ Arbitrage-free pricing of Credit Index Options.\\
The no-armageddon pricing measure and the role of correlation after the
subprime crisis\bigskip }
\author{Massimo Morini\thanks{%
Corresponding Author. We thank Tomasz Bielecki, Paolo Longato and Lutz Schloegl for helpful
comments and discussion. Tomasz Bielecki also signalled us that a related work on credit index options is being developed independently in Armstrong and Rutkowski (2007) \newline
\newline
This paper expresses the views of its authors and does not represent the
opinion of Banca IMI or FitchRatings, and neither organization is
responsible for any use which maybe made of its contents.}\medskip \\
{\small Banca IMI, Intesa-SanPaolo, and}\\
{\small Dept. of Quant. Methods, Bocconi University.}\\
\and Damiano Brigo\medskip \\
{\small Fitch Solutions, 101 Finsbury Pavement, London}\\
{\small and Dept. of Mathematics, Imperial College.}}
\date{\medskip \medskip \medskip First Version: September 2007. This Version:
1 December 2007}
\maketitle

\pagestyle{myheadings} \markboth{}{{\footnotesize  Massimo Morini and Damiano Brigo - Arbitrage Free Pricing of Credit Index Options}}

\begin{abstract}
In this work we consider three problems of the standard market approach to
pricing of credit index options: the definition of the index spread is not
valid in general, the usually considered payoff leads to a pricing which is
not always defined, and the candidate numeraire one would use to define a
pricing measure is not strictly positive, which would lead to a
non-equivalent pricing measure.

We give a general mathematical solution to the three problems, based on a
novel way of modelling the flow of information through the definition of a
new subfiltration. Using this subfiltration, we take into account
consistently the possibility of default of all names in the portfolio, that
is neglected in the standard market approach. We show that, while the
related mispricing can be negligible for standard options in normal market
conditions, it can become highly relevant for different options or in
stressed market conditions.

In particular, we show on 2007 market data that after the subprime credit
crisis the mispricing of the market formula compared to the no arbitrage
formula we propose has become financially relevant even for the liquid
Crossover Index Options. \newpage
\end{abstract}

\tableofcontents

\newpage

\section{Introduction}

In this paper we address the issue of the correct pricing of options on
credit swaps, where the reference entity is a portfolio of defaultable
names. The most common case in the market are Credit Index Options, which
are options on the equilibrium level of the spread of a Credit Index,
consisting of a standardized portfolio of credit default swaps. Liquidity
had improved in the years preceding the subprime crisis in summer 2007, in
particular in Europe. Monthly volumes for the option on the i-Traxx
Crossover Index were upwards 13 billion at the end of 2006, with options
available also on the Main and High Volatility Indices and on Index
Tranches. During the credit crisis in 2007, payer options have protected
investors from the rise of the credit indices, and trading has restarted on
these products just two weeks after the pitch of the crisis at the end of
July 2007.

Thus the main reference in the paper will be Credit Index swaptions,
although the results apply in general to all market cases in which a spread
defined with reference to a portfolio of defaultable entities must be
modelled in an arbitrage-free way under an appropriate pricing measure. This
includes dynamic notes linked to the performances of Index Spreads and
embedded optionality in multiname credit derivatives.

The Credit Index Option gives to the investor the possibility to enter a
Forward Credit Index at a prespecified spread, and to receive upon exercise
a Front End Protection corresponding to index losses from option inception
to option expiry. The initial market approach for pricing Credit Index
Options was the use of a Black formula to price the option as a call on the
spread, adding then the value of the Front End Protection to the price of
the call. This approach has a first major flaw. It neglects the fact that
one cannot separate the option from the Front End Protection, since the
protection enters the exercise decision. This was pointed out first by
Pedersen (2003). This observation led to an improved market standard, taking
Front End Protection into account correctly without losing the tractability
of the market Black formula approach, based on a redefinition of the
underlying Index spread.

However, as we show in this work, also the latter approach presents some
relevant problems.

First, there is still one scenario in which a Black formula, even if
improved by the use of a Loss-adjusted Index Spread, does not take into
account correctly the Front End Protection. In this scenario the pricing
formula does not give the correct price of the option, while instead this
could be computed in closed-form. We show in this work that, while this
mispricing can be negligible for standard options in normal market
conditions, it can become highly relevant for less standard options or in
stressed market conditions. In particular, we show on 2007 market data that
after the subprime credit crisis the mispricing of the market formula
compared to the no arbitrage formula we propose has become financially
relevant even for the most liquid Crossover Index Options.

A related second problem is that the formula used in market practice to
compute the Index Spread is not defined in all states of the world.

The third problem regards the theoretical justification of the use of a
Black formula in this context. According to the Fundamental Theorem of Asset
Pricing, a rigorous derivation of a Black formula for the pricing of a
Credit Index Option requires the definition of an appropriate numeraire for
change to the pricing measure under which the underlying spread is a
martingale (see Delbaen and Schachermayer (1994), Geman et al. (1995), and
Jamshidian (1997)). The problem here is that, since we are dealing with
portfolios or indices of defaultable names, the quantity that appears to be
the natural choice for a numeraire to simplify pricing is not strictly
positive. This makes the standard change of numeraire theory, based on
strictly positive numeraires, inapplicable.

One may use a kind of survival measure such as the one introduced by
Schonbucher (1999) for single name products, based on a numeraire which is
not strictly positive. But such a measure would not be equivalent to the
standard risk neutral, forward and swap measures used in mathematical
finance, but only absolutely continuous with respect to them. This is
practically undesirable, since in a such a case one can describe the
dynamics of credit spreads under the survival measure, but cannot use the
standard Girsanov Theorem to see the dynamics of credit spreads under the
standard measures used in mathematical finance. Thus one could not, for
example, extend a standard implementation of a Libor Market Model (Brace,
Gatarek and Musiela (1997)) for default-free forward rates, usually done
under a forward measure, to include also forward credit spreads.

To the best of our knowledge, the current literature does not solve the
above three problems. The only partial exception is Jackson (2005), that
deals with the first problem while not solving the second and third ones (in
particular he uses a numeraire which is not strictly positive). Instead,
here we show that the definition of an index spread valid in general, the
correct description of the market payoff leading to a price defined in all
states of the world, and the use of a valid strictly positive quantity to
define an equivalent pricing measure can be given a general and elegant
mathematical solution. The solution is based on the definition of
appropriate subfiltrations. Empirical tests show that this solution can also
have a relevant impact on the correct price of multiname credit options.

In Section \ref{Section_Themarket} we introduce the setting and describe the
Credit Forward Index, that will be the reference underlying of the options
considered in the paper. Then we describe the Credit Index Option Payoff and
we describe previous literature and market practice on the pricing of Index
options. In Section \ref{Section_IndexSpread} we show the problems in the
market approach and we introduce the main technical instrument that will
allow us to solve them: subfiltrations. In particular we introduce a new
subfiltration apt to credit portfolio products. Then we use this instrument
to compute a consistent and arbitrage-free definition of the underlying
Index spread. In Section \ref{Section_IndexOption} we introduce a new
pricing measure and prove the main result of the paper, the formula for
no-arbitrage pricing of Index Options, different from the standard market
approach. We also show how, performing generalization of some fundamental
results of Jamshidian (2004), it is possible to express this formula in
terms of numeraire pricing. In Section \ref{Section_Tests} we apply our
results to market data, before and after the summer 2007 subprime crisis,
and we show how, in stressed market conditions like those the credit market
is experiencing now, the no-arbitrage formula we introduce has a relevant
impact on valuation of Credit Index Options, avoiding a mispricing which is
associated to the market standard valuation formula.

\section{The Market\label{Section_Themarket}}

We are in a complete filtered probability space $\left( \Omega ,\mathcal{F}%
,P,\mathcal{F}_{t}\right) $, where the filtration $\mathcal{F}_{t}$
satisfies the usual hypothesis and we set $\mathcal{F}_{0}=\left( \Omega
,\emptyset \right) $ and $\mathcal{F=F}_{\bar{T}}$ for a terminal date $0<%
\bar{T}<\infty $. We assume the existence of a bank-account numeraire with
price process $B_{t}$ and an associated risk neutral measure $\mathbb{Q}\sim
P$. Defining, in line with Jamshidian (2004), a claim $X=X_{\bar{T}}$ as an $%
\mathcal{F}$-measurable random variable such that $\frac{X}{B_{\bar{T}}}$ is
$\mathbb{Q}$-integrable, the price process of any claim is given by
\begin{equation*}
X_{t}=\mathbb{E}\left[ D\left( t,\bar{T}\right) X\mathbf{|}\mathcal{F}_{t}%
\right]
\end{equation*}
where $\mathbb{E}$ indicates expectation under the risk-neutral measure and
\begin{equation*}
D\left( t,T\right) =\frac{B_{t}}{B_{T}}
\end{equation*}
is the discount factor from $T$ to $t$. We consider $n$ defaultable issuers
in a portfolio, and we indicate by $\tau _{i}$ the time of default of the $i$%
-th issuer, $i=1,\ldots ,n$.

In the following we are going to use Credit Indices as a reference. However
our results can be extended to different multiname credit swaps (such as
tranches) mainly through a redefinition of the loss from which protection is
sought. The Credit Index is a swap contract providing protection against the
default of a pool of $n$ names, from the beginning of the contract $T_{A}$
and until maturity $T_{M}$. A spread or premium $K$ is paid for the Index
protection, proportional to an \emph{outstanding notional} that diminishes
as names in the pool default.

There are two main families of indices. The \emph{Dow Jones i-Traxx Indices}
represent (mainly) the European market. The \emph{CDX} family represent the
North American market. In particular, for Main Indices the portfolio is set
up to include only investment grade names (in fact the Main Indices are also
called \emph{Investment Grade Indices}), while the Crossover Indices include
names with a lower credit quality. In the CDX family there is also a \emph{%
High Yield Index} usually including issuers with credit reliability lower
than Crossover names.

\subsection{The Forward Credit Index\label{Section_forwardIndex}}

For a unit of Index notional each name has a notional of $\frac{1}{n}$. It
is market standard to give index quotations assuming that also the \emph{%
recovery rate} ($R$) is the same for all names. The cumulated loss at $t$ is
\begin{equation*}
L\left( t\right) =\frac{\left( 1-R\right) }{n}\sum_{i=1}^{n}1_{\left\{ \tau
_{i}\leq t\right\} }
\end{equation*}
At time $t$ the Outstanding Notional is
\begin{equation*}
N\left( t\right) =1-\frac{L\left( t\right) }{\left( 1-R\right) }.
\end{equation*}
In a Credit Index the \emph{Protection Leg} pays, at default times $\tau
_{i} $, the corresponding loss $dL\left( \tau _{i}\right) $, from start date
$T_{A}$ to maturity $T_{M}$ or until all names have defaulted. The \emph{%
Premium Leg} pays, at times $T_{j}$, $j=A+1,\ldots ,M$ or until all names
have defaulted, a premium $K$ on the average $\hat{N}\left( T_{j}\right) $
of the outstanding notional $N\left( t\right) $ for $t\in \left(
T_{j-1},T_{j}\right] $. The discounted payoff of the Protection Leg is
\begin{eqnarray}
\Phi _{t}^{T_{A},T_{M}} &=&\int_{T_{A}}^{T_{M}}D\left( t,u\right) dL\left(
u\right)  \label{formula_protectionleg} \\
&\approx &\sum_{j=A+1}^{M}D\left( t,T_{j}\right) \left[ L\left( T_{j}\right)
-L\left( T_{j-1}\right) \right] ,  \notag
\end{eqnarray}
where the approximation discretizes loss payment times to the standard
payment times of the Premium Leg.

The discounted payoff of the Premium Leg is
\begin{eqnarray}
&&\Psi _{t}^{T_{A},T_{M}}\left( K\right)
\begin{tabular}{l}
$=$%
\end{tabular}
\;\left\{ \sum_{j=A+1}^{M}D\left( t,T_{j}\right)
\int_{T_{j-1}}^{T_{j}}N\left( t\right) dt\right\} K
\label{formula_premiumleg} \\
&&
\begin{tabular}{l}
$\approx $%
\end{tabular}
\;\left\{ \sum_{j=A+1}^{M}D\left( t,T_{j}\right) \alpha _{j}\left( 1-\frac{%
L\left( T_{j}\right) }{\left( 1-R\right) }\right) \right\} K,  \notag
\end{eqnarray}
where the discretization considers the outstanding notional as computed at
the end of the interval $\left( T_{j-1},T_{j}\right] $. The interval has
length $\alpha _{j}$. The value of the two legs is computed by expectation
under the risk neutral probability measure $\mathbb{Q}$. We use the notation
$\Pi \left( X_{t},s\right) =\mathbb{E}\left[ \left. X_{t}\right| \mathcal{F}%
_{s}\right] $. When $s=t$, we omit the argument $s$ writing simply $\Pi
\left( X_{t}\right) $. For the value of the two legs we have
\begin{equation*}
\Pi \left( \Phi _{t}^{T_{A},T_{M}}\right) :=\mathbb{E}\left[ \left. \Phi
_{t}^{T_{A},T_{M}}\right| \mathcal{F}_{t}\right] ,\;\;\;\;\Pi \left( \Psi
_{t}^{T_{A},T_{M}}\left( K\right) \right) :=\mathbb{E}\left[ \left. \Psi
_{t}^{T_{A},T_{M}}\left( K\right) \right| \mathcal{F}_{t}\right]
\end{equation*}
The quantity in curly brackets in (\ref{formula_premiumleg}) is called in
market practice the \emph{Index Defaultable present Value per Basis Point},
\begin{equation*}
\Pi \left( \gamma _{t}^{T_{A},T_{M}}\right) =:\mathbb{E}\left[ \left. \gamma
_{t}^{T_{A},T_{M}}\right| \mathcal{F}_{t}\right] :=\mathbb{E}\left[ \left.
\sum_{j=A+1}^{M}D\left( t,T_{j}\right) \alpha _{j}\left( 1-\frac{L\left(
T_{j}\right) }{\left( 1-R\right) }\right) \right| \mathcal{F}_{t}\right]
\end{equation*}
The Payer Forward Index starting at $T_{A}$ and lasting until $T_{M}$ has a
payoff discounted at $t$ given by
\begin{equation*}
I_{t}^{T_{A},T_{M}}\left( K\right) =\Phi _{t}^{T_{A},T_{M}}-\Psi
_{t}^{T_{A},T_{M}}\left( K\right) .
\end{equation*}
while its price is given by
\begin{equation*}
\Pi \left( I_{t}^{T_{A},T_{M}}\left( K\right) \right) :=\mathbb{E}\left[
\left. I_{t}^{T_{A},T_{M}}\left( K\right) \right| \mathcal{F}_{t}\right] :=%
\mathbb{E}\left[ \left. \Phi _{t}^{T_{A},T_{M}}-\Psi
_{t}^{T_{A},T_{M}}\left( K\right) \right| \mathcal{F}_{t}\right]
\end{equation*}

We have introduced forward index quantities because the Forward Index is the
underlying of the most common credit portfolio option, the Index Swaption.
It is useful to analyze these quantities in terms of single name defaults in
order to understand how forward index quotations are made available in the
market. In terms of single name quantities, the Protection Leg has value
\begin{equation*}
\Pi \left( \Phi _{t}^{T_{A},T_{M}}\right) =\sum_{j=A+1}^{M}\sum_{i=1}^{n}%
\frac{\left( 1-R\right) }{n}\mathbb{E}\left[ \left. D\left( t,T_{j}\right)
\left( 1_{\left\{ T_{j-1}<\tau _{i}\leq T_{j}\right\} }\right) \right|
\mathcal{F}_{t}\right] ,
\end{equation*}
while the Premium Leg has value
\begin{equation*}
\Pi \left( \Psi _{t}^{T_{A},T_{M}}\left( K\right) \right)
=\sum_{j=A+1}^{M}\sum_{i=1}^{n}\frac{1}{n}\alpha _{j}\mathbb{E}\left[ \left.
D\left( t,T_{j}\right) 1_{\left\{ \tau _{i}>T_{j}\right\} }\right| \mathcal{F%
}_{t}\right] K.
\end{equation*}

Forward Index quotations $\Pi \left( I_{t}^{T_{A},T_{M}}\left( K\right)
\right) $ are not directly provided by the market, since only spot indices
are quoted, namely $\Pi \left( I_{0}^{0,T_{M}}\left( K\right) \right) $ with
$T_{M}\in \mathbb{T}$, where $\mathbb{T}$ is a standardized set of annual
maturities. Forward quotations are extracted from spot quotations according
to the following homogeneity modelling assumptions. The first simplification
for quoting indices is to assume that interest rates are independent of
default, so
\begin{eqnarray*}
\Pi \left( \Phi _{t}^{T_{A},T_{M}}\right) &=&\sum_{j=A+1}^{M}\frac{1}{n}%
\left( 1-R\right) \sum_{i=1}^{n}P\left( t,T_{j}\right) \mathbb{Q}\left(
\left. T_{j-1}<\tau _{i}\leq T_{j}\right| \mathcal{F}_{t}\right) , \\
\Pi \left( \Psi _{t}^{T_{A},T_{M}}\left( K\right) \right) &=&\sum_{j=A+1}^{M}%
\frac{1}{n}\sum_{i=1}^{n}\alpha _{j}P\left( t,T_{j}\right) \mathbb{Q}\left(
\left. \tau _{i}>T_{j}\right| \mathcal{F}_{t}\right) \times K,
\end{eqnarray*}
where $P\left( t,T\right) $ is the price at $t$ of the default-free bond
with maturity $T$. The second simplification introduced for quoting indices
is the assumption of \emph{homogeneous portfolio}\textbf{,} corresponding to
assuming that all names have the same credit risk, and consequently the same
survival and default probability. This default probability common to all
names is expressed through intensity modelling (Lando (1998)), with a
deterministic, time-dependent intensity $\lambda _{s}$. One finally obtains
\begin{equation}
\Pi \left( \Phi _{t}^{T_{A},T_{M}}\right) =N\left( t\right) \left(
1-R\right) \sum_{j=A+1}^{M}P\left( t,T_{j}\right) \left(
e^{-\int_{t}^{T_{j-1}}\lambda _{s}ds}-e^{-\int_{t}^{T_{j}}\lambda
_{s}ds}\right) ,  \label{formula_protLeg_Intensities}
\end{equation}
and
\begin{equation}
\Pi \left( \Psi _{t}^{T_{A},T_{M}}\left( K\right) \right) =N\left( t\right)
\sum_{j=A+1}^{M}\alpha _{j}P\left( t,T_{j}\right) \times
e^{-\int_{t}^{T_{j}}\lambda _{s}ds}\times K  \label{formula_DV01_intensities}
\end{equation}
The default intensity is obtained through calibration to Spot Indices, with $%
T_{A}=0$, and then used to compute quotations for forward indices, providing
the underlying of Credit Index options. Credit Index options are introduced
in the next section. In treating Index Options we will not confine ourselves
to the simplifying hypotheses used by the market standard for index
quotations.

\subsection{The Credit Index Option \label%
{Section_Index_Option_payoff_and Standard}}

A\textbf{\ }\emph{payer Credit Index Option} with inception $0$, strike $K$
and exercise date $T_{A}$, written on an index with maturity $T_{M}$, is a
contract giving the right to enter at $T_{A}$\ into an Index with final
payment at $T_{M}$, as a \emph{protection buyer}\ paying a fixed rate $K$,
thus entitled to receive protection from losses in the period between $T_{A}$
and $T_{M}$. In addition the protection buyer receives, upon exercise, also
the so-called \emph{Front End Protection}, covering the losses from the
option inception at time $0$ to the exercise date $T_{A}.$ The Front End
Protection and its financial rationale are described below.

The purpose of a Credit Index Option is to allow the protection buyer to
lock in a particular premium spread $K$, that the protection buyer has the
right (and not the obligation) to make effective at a future time. However,
if the above simple payoff was considered without Front End Protection,
there would be an additional side-effect: the option buyer would not be
protected from losses in the period between $0$ and maturity $T_{A}$. In
order to attract more investors, the standard Credit Index Option payoff
includes, in case the option is exercised, the Front End Protection, whose
discounted payoff $F_{t}^{T_{A}}$ is given by
\begin{eqnarray*}
F_{t}^{T_{A}} &=&D\left( t,T_{A}\right) L\left( T_{A}\right) \\
\Pi \left( F_{t}^{T_{A}}\right) &=&\mathbb{E}\left[ \left. D\left(
t,T_{A}\right) L\left( T_{A}\right) \right| \mathcal{F}_{t}\right] .
\end{eqnarray*}
In this section we present the market approach to the pricing of Credit
Index Options, and then in the next sections we point out the open problems
and present our solutions. In the simplest market approach, the Index Option
is priced as a call option on the Index spread, and then the value of the
front-end protection is added to the option price. The Index Spread is
defined as the value of $K$ that sets the index value $\Pi \left(
I_{t}^{T_{A},T_{M}}\left( K\right) \right) $ to zero,
\begin{equation*}
S_{t}^{T_{A},T_{M}}=\frac{\Pi \left( \Phi _{t}^{T_{A},T_{M}}\right) }{\Pi
\left( \gamma _{t}^{T_{A},T_{M}}\right) }.
\end{equation*}
allowing to write the index value as
\begin{equation*}
\Pi \left( I_{t}^{T_{A},T_{M}}\left( K\right) \right) =\Pi \left( \gamma
_{t}^{T_{A},T_{M}}\right) \left( S_{t}^{T_{A},T_{M}}-K\right)
\end{equation*}
The Index option is then priced by decomposing its value as follows
\begin{equation*}
\mathbb{E}\left[ \left. D\left( t,T_{A}\right) \Pi \left( \gamma
_{T_{A}}^{T_{A},T_{M}}\right) \left( S_{T_{A}}^{T_{A},T_{M}}-K\right)
^{+}\right| \mathcal{F}_{t}\right] \mathbb{+}\Pi \left( F_{t}^{T_{A}}\right)
\end{equation*}
and then by expressing the first component through a standard Black formula,
\begin{equation}
\Pi \left( \gamma _{t}^{T_{A},T_{M}}\right) Black\left(
S_{t}^{T_{A},T_{M}},K,\sigma ^{T_{A},T_{M}}\sqrt{T_{A}-t}\right) \mathbb{+}%
\Pi \left( F_{t}^{T_{A}}\right)  \label{formula_wrong1_option_price}
\end{equation}
where $\sigma ^{T_{A},T_{M}}$ is the volatility of the forward spread and
\begin{eqnarray*}
Black\left( S,K,\sigma \right) &=&S\mathbb{\times }\mathbb{N}\left(
d_{1}\right) -K\mathbb{\times }\mathbb{N}\left( d_{2}\right) \\
d_{1} &=&\frac{\ln \left( \frac{S}{K}\right) +\frac{1}{2}\sigma ^{2}}{\sigma
},\;\;d_{2}=\frac{\ln \left( \frac{S}{K}\right) -\frac{1}{2}\sigma ^{2}}{%
\sigma }.
\end{eqnarray*}

This standard market approach to pricing the index option has a main flaw,
as noticed by Pedersen (2003). The above formula neglects the fact that the
front end protection is received only upon exercise. Pedersen (2003),
defines the option payoff as
\begin{equation}
D\left( t,T_{A}\right) \left( \Pi \left( \gamma
_{T_{A}}^{T_{A},T_{M}}\right) \left( S_{T_{A}}^{T_{A},T_{M}}-K\right)
+F_{T_{A}}^{T_{A}}\right) ^{+}  \label{formula_wrong2_option_payoff}
\end{equation}
and suggests a numerical integration to compute the expectation of payoff (%
\ref{formula_wrong2_option_payoff}). The prices of Index swaptions computed
with Pedersen's Formula confirm that the standard approach overprices
out-of-the money payer options. Pedersen (2003) results have led to an
improved Black Formula approach, based on the redefinition of the underlying
Index Spread.

The procedure is described for example in Doctor and Goulden (2007). One
defines a \emph{Loss-Adjusted }index spread by taking Front End Protection
into account directly. In fact the actual underlying of the credit index
option is the \emph{Loss-Adjusted Index},
\begin{equation}
\tilde{I}_{t}^{T_{A},T_{M}}\left( K\right) =\Phi _{t}^{T_{A},T_{M}}-\Psi
_{t}^{T_{A},T_{M}}\left( K\right) +F_{t}^{T_{A}}.
\label{formula_loss_adjusted_Index_payoff}
\end{equation}
Thanks to the market standard quotation system, seen in Section \ref
{Section_forwardIndex}, one has market liquid information available to
compute all the above quantities. So it is natural to give a new spread
definition, setting to zero $\Pi \left( \tilde{I}_{t}^{T_{A},T_{M}}\left(
K\right) \right) $ rather than $\Pi \left( I_{t}^{T_{A},T_{M}}\left(
K\right) \right) $. This leads to the following \emph{Loss-Adjusted Market
Index Spread}
\begin{equation}
\tilde{S}_{t}^{T_{A},T_{M}}=\frac{\Pi \left( \Phi _{t}^{T_{A},T_{M}}\right)
+\Pi \left( F_{t}^{T_{A}}\right) }{\Pi \left( \gamma
_{t}^{T_{A},T_{M}}\right) }  \label{formula_market_index_spread}
\end{equation}
that allows to write (\ref{formula_wrong2_option_payoff}) as
\begin{equation}
D\left( t,T_{A}\right) \left( \Pi \left( \gamma
_{T_{A}}^{T_{A},T_{M}}\right) \left( \tilde{S}_{t}^{T_{A},T_{M}}-K\right)
\right) ^{+}.  \label{formula_wrong3_option_payoff}
\end{equation}
Now one can think of taking $\Pi \left( \gamma _{t}^{T_{A},T_{M}}\right) $
as numeraire and $\tilde{S}_{t}^{T_{A},T_{M}}$ as lognormal underlying
variable so as to price the option with the \emph{Market Credit Index Option
Formula}:
\begin{equation}
\Pi \left( \gamma _{t}^{T_{A},T_{M}}\right) Black\left( \tilde{S}%
_{t}^{T_{A},T_{M}},K,\tilde{\sigma}^{T_{A},T_{M}}\sqrt{T_{A}}\right) .
\label{formula_wrong2_option_price}
\end{equation}
A\textbf{\ }\emph{receiver Credit Index Option} with the same contract
specifications as the payer above, is a contract giving the right to enter
at $T_{A}$\ into the same Index with final payment at $T_{M}$, as a \emph{%
protection seller}\ receiving a fixed rate $K$. Following the same steps as
above, the payoff is
\begin{equation*}
D\left( t,T_{A}\right) \left( \Pi \left( \gamma
_{T_{A}}^{T_{A},T_{M}}\right) \left( K-S_{T_{A}}^{T_{A},T_{M}}\right)
-F_{T_{A}}^{T_{A}}\right) ^{+}
\end{equation*}
and the price is given by
\begin{equation*}
\Pi \left( \gamma _{t}^{T_{A},T_{M}}\right) \times \left( K\mathbb{\times }%
\mathbb{N}\left( -d_{2}\right) -\tilde{S}_{t}^{T_{A},T_{M}}\mathbb{\times }%
\mathbb{N}\left( -d_{1}\right) \right) .
\end{equation*}

\section{Towards an arbitrage-free Index Spread\label{Section_IndexSpread}}

We have seen in the previous Section that it appears possible to take Front
End Protection into account correctly without losing the tractability of the
market Black formula approach.

This is an important issue for building a solid and standardized option
market. A Black formula approach is for example the standard for pricing
swap options in the interest rate world, as justified by the market model
framework introduced by Jamshidian (1997) for swaptions and Brace, Gatarek
and Musiela (1997) for caps. In this pricing approach, the complexity of the
no-arbitrage dynamics of the underlying asset is consistently transferred to
a correct numeraire, through change of measure.

However, the market option formula (\ref{formula_wrong2_option_price}) does
not represent yet a consistent extension of this approach to Credit Index
Options. If one wants to reach a consistent definition of the Index spread
and a safely arbitrage-free valuation formula, one needs a specific
treatment of the information sets involved, which in presence of default
portfolios is more subtle than in the default-free case or even the
single-name default case. This is shown in the following.

\subsection{Problems of the market formulas\label%
{Section_advantages&problems}}

The approach leading to the market option formula (\ref
{formula_wrong2_option_price}) has highly improved on the initial market
approach leading to (\ref{formula_wrong1_option_price}). However, even the
most recent approach fails. The two approaches share the following problems.

\begin{enumerate}
\item  The definition of the spread $\tilde{S}_{t}^{T_{A},T_{M}}$ is not
valid globally, but only when the denominator
\begin{equation*}
\Pi \left( \gamma _{t}^{T_{A},T_{M}}\right) =\sum_{j=A+1}^{M}\mathbb{E}\left[
\left. D\left( t,T_{j}\right) \alpha _{j}\left( 1-\frac{L\left( T_{j}\right)
}{\left( 1-R\right) }\right) \right| \mathcal{F}_{t}\right]
\end{equation*}
is different from zero. Since $\Pi \left( \gamma _{t}^{T_{A},T_{M}}\right) $
is the price of a portfolio of defaultable assets, this quantity may vanish,
it is not bounded away from zero in all states of the world having positive
probability.

\item  When $\Pi \left( \gamma _{t}^{T_{A},T_{M}}\right) =0$ the pricing
formula (\ref{formula_wrong2_option_price}) is undefined, while instead we
will see that the option price is known exactly in such a scenario. We will
see, additionally, that in this scenario the above spread $\tilde{S}%
_{t}^{T_{A},T_{M}}$ does not set the value of the adjusted index to zero.

\item  Since it is not strictly positive, $\Pi \left( \gamma
_{t}^{T_{A},T_{M}}\right) $ would lead to the definition of a pricing
measure not equivalent to the standard risk-neutral measure.
\end{enumerate}

To the best of our knowledge, the current literature does not solve these
problems. The only partial exception is Jackson (2005), that deals with 2)
while not considering 1) and 3). Instead, here we give a solution to the
above problems based on the definition of appropriate subfiltrations.

\subsection{Subfiltrations for single issuer}

Problems similar to those presented in the last section appear also in
modelling default of a single name for pricing an option on a Credit Default
Swap (CDS). A CDS is like an index swap with one single name. The CDS price
for protection from $T_{a}$ and $T_{b}$, in a simple formulation neglecting
the accrual terms, is
\begin{equation}
\Pi \left( CDS_{t}^{T_{a},T_{b}}\left( K\right) \right)
\begin{tabular}{l}
$=$%
\end{tabular}
\left( 1-R\right) \sum_{i=a+1}^{b}\mathbb{E}^{\mathbb{Q}}\left[ D\left(
t,T_{i}\right) \mathbf{1}_{\left\{ T_{i-1}<\tau \leq T_{i}\right\} }\mathbf{|%
}\mathcal{F}_{t}\right] -K\sum_{i=a+1}^{b}\mathbb{E}^{\mathbb{Q}}\left[
D(t,T_{i})\alpha _{i}\mathbf{1}_{\left\{ \tau >T_{i}\right\} }\mathbf{|}%
\mathcal{F}_{t}\right] ,
\end{equation}
where $\tau $ indicates the default time of the single reference credit
entity.

When faced with an option on a CDS, it is natural to define the CDS
equilibrium spread $K_{t}^{T_{a},T_{b}}$ as the level of $K$ setting the
above price to zero, and to chose as a numeraire for change of measure the
single name defaultable present value per basis point
\begin{equation*}
\sum_{i=a+1}^{b}\mathbb{E}^{\mathbb{Q}}\left[ D(t,T_{i})\alpha _{i}\mathbf{1}%
_{\left\{ \tau >T_{i}\right\} }\mathbf{|}\mathcal{F}_{t}\right]
\end{equation*}
However, this is not a viable numeraire since it has zero value in all
states of the world where the underlying name defaults before the option
maturity, a set with positive measure. Thus, in all such states of the world
the above definition of the CDS equilibrium spread $K_{t}^{T_{a},T_{b}}$ is
not valid. These problems of single name credit modelling are dealt with by
Jamshidian (2004) and Brigo (2005, 2008). Although the solution proposed in these
works cannot be trivially extended to the multiname setting, we recall them
in the following since they are the foundations of the results we present
here.

Jamshidian (2004) and Brigo (2005, 2008) make use of a \emph{subfiltration
structure}. Following Jeanblanc and Rutkowski (2000), define $\mathcal{F}%
_{t}=\mathcal{J}_{t}\vee \mathcal{H}_{t}$, where
\begin{equation*}
\mathcal{J}_{t}=\sigma \left( \left\{ \tau >u\right\} ,u\leq t\right) ,
\end{equation*}
namely the filtration generated by $\tau $, thus representing default
monitoring up to $t$, while $\mathcal{H}_{t}$ is a filtration representing
the flow of all information except default itself (default-free
information). A market operator observing only this second filtration can
have information on the probability of default but cannot say exactly when,
or even if, default has happened. This structure is typical for instance of
the Cox process setting, where default is defined as the first jump of a Cox
Process. A Cox Process is a process that, conditional on the path $\left(
\lambda _{t}\right) _{t\geq 0}$ followed by the stochastic default intensity
$\lambda _{t}$, is a Poisson Process of intensity $\left( \lambda
_{t}\right) _{t\geq 0}$. This definition hinges on assuming default
intensity $\lambda _{t}$ to be $\mathcal{H}_{t}$-adapted.\ However, the use
of subfiltrations is not limited to Cox Processes.

The advantage of subfiltrations in the context of credit option pricing is
that a subfiltration structure allows to define pricing formulas in terms of
conditional survival probability $\mathbb{Q}\left( \tau >t|\mathcal{H}%
_{t}\right) $ which can be assumed to be strictly positive in any state of
the world (or at least a.s.). In particular it allows to use a result by
Jeanblanc and Rutkowski (2000), based on the definition of conditional
expectation. A defaultable payoff with maturity $T$ discounted to $t$, $%
\mathbf{Y}_{t}^{T}$, is a payoff that satisfies
\begin{equation}
\;\;\;\;\;\;\;\;\;\;\;\;\;\;\;\;\;\;\;\;\;\;\;\;\;\;\;\;\;\;\;\;\;\;\;\;\;\;%
\;\;\;\;\;\;\;\;\;\;\;\;\;\;\;\;\;\;\;\;\;\;\;\;\;\;\;\mathbf{Y}_{t}^{T}=%
\mathbf{1}_{\left\{ \tau >T\right\} }\mathbf{Y}_{t}^{T},
\label{formula_definition_of_defaultable}
\end{equation}
allowing us to write the \emph{Jeanblanc and Rutkowski Formula}
\begin{equation}
\Pi \mathbf{Y}_{t}^{T}=\mathbb{E}\left[ \mathbf{Y}_{t}^{T}\mathbf{|}\mathcal{%
F}_{t}\right] =\frac{\mathbf{1}_{\left\{ \tau >t\right\} }}{\mathbb{Q}\left(
\tau >t|\mathcal{H}_{t}\right) }\mathbb{E}\left[ \mathbf{Y}_{t}^{T}|\mathcal{%
H}_{t}\right] .  \label{formula_JeanRutk_defaultable}
\end{equation}
Since a CDS is a defaultable contract, $CDS_{t}^{T_{a},T_{b}}\left( K\right)
=\mathbf{1}_{\left\{ \tau >T_{a}\right\} }CDS_{t}^{T_{a},T_{b}}\left(
K\right) $, one can write
\begin{equation}
\Pi \left( CDS_{t}^{T_{a},T_{b}}\left( K\right) \right) =\frac{\mathbf{1}%
_{\left\{ \tau >t\right\} }}{\mathbb{Q}\left( \tau >t|\mathcal{H}_{t}\right)
}\mathbb{E}^{\mathbb{Q}}\left[ CDS_{t}^{T_{a},T_{b}}\left( K\right) |%
\mathcal{H}_{t}\right] .  \label{subfiltration pricing}
\end{equation}
Setting expression (\ref{subfiltration pricing}) to zero and solving in $K$
one find a definition of $K_{t}^{T_{a},T_{b}}$ with a regular behaviour in
all scenarios
\begin{equation*}
K_{t}^{T_{a},T_{b}}=\left( 1-R\right) \frac{\sum_{i=a+1}^{b}\mathbb{E}^{%
\mathbb{Q}}\left[ D\left( t,T_{i}\right) \mathbf{1}_{\left\{ T_{i-1}<\tau
\leq T_{i}\right\} }|\mathcal{H}_{t}\right] }{\sum_{i=a+1}^{b}\alpha _{i}%
\mathbb{E}^{\mathbb{Q}}\left[ D\left( t,T_{i}\right) \mathbf{1}_{\left\{
\tau >T_{i}\right\} }|\mathcal{H}_{t}\right] }
\end{equation*}
since the denominator, being a.s. strictly positive, will also allow to
define a probability measure that, under a Cox Process setting, leads to a
simple Black formula, as shown in Brigo (2005, 2008):
\begin{equation*}
\frac{\mathbf{1}_{\left\{ \tau >t\right\} }}{\mathbb{Q}\left( \tau >t|%
\mathcal{H}_{t}\right) }\left\{ \sum_{i=a+1}^{b}\mathbb{E}^{\mathbb{Q}}\left[
D\left( t,T_{i}\right) \alpha _{i}\mathbf{1}_{\left\{ \tau >T_{i}\right\} }|%
\mathcal{H}_{t}\right] \right\} Black\left( K_{t}^{T_{a},T_{b}},K,\bar{\sigma%
}_{a,b}\sqrt{T_{a}-t}\right) .
\end{equation*}
In order to understand the following developments, it is important to notice
why the use of the $\mathcal{H}_{t}$ subfiltration, excluding information on
default of the underlying single name, has effectively allowed to solve the
above problems in a single name context. The corporate zero-coupon bonds
considered in the CDS numeraire definition and the CDS itself are
defaultable payoffs, whose value goes to zero at default, so we can use (\ref
{formula_JeanRutk_defaultable}) and we can confine ourselves to making
assumptions on the stochastic dynamics of the credit spread only on the $%
\mathcal{H}_{t}$ subfiltration, which is all we need for pricing when $%
\left\{ \tau >t\right\} $. We do not need to treat explicitly the case $%
\left\{ \tau \leq t\right\} $ since all involved payoffs have in this case a
value which is known to be zero.

Then, for applying these results to options, we need the option itself to be
priced with (\ref{formula_JeanRutk_defaultable}), and this is possible
because single name Credit Options are \emph{knock-out options}, namely
their value is zero after default.

The situation in a multiname setting is different, so it requires a related
but different approach, outlined in the next section.

\subsection{A new subfiltration for multiname credit derivatives}

In a multiname setting we have a plurality of reference entities that can
default, so the above subfiltration setting gives us a plurality of possible
subfiltrations,
\begin{eqnarray}
\mathcal{F}_{t} &=&\mathcal{J}_{t}^{i}\vee \mathcal{H}_{t}^{i},
\label{formula_Jamshidian_information_structure} \\
\mathcal{J}_{t}^{i} &=&\sigma \left( \left\{ \tau _{i}>u\right\} ,u\leq
t\right) ,  \notag
\end{eqnarray}
so that $\mathcal{J}_{t}^{i}$ is the filtration generated by the default
time $\tau _{i}$ of the name $i$, with $i=1,2,\ldots ,n,$ while $\mathcal{H}%
_{t}^{i}$ is a filtration representing the flow of all information except\
the default of the name $i$. When the information structure of a single name
default valuation model is defined implicitly through (\ref
{formula_Jamshidian_information_structure}), Jamshidian (2004) calls the
resulting model \emph{complementary}. Complementary models are central in
the framework of Jamshidian (2004).

It would be enough to consider any of these possible subfiltrations $%
\mathcal{H}_{t}^{i}$ to avoid the multiname Index spread $\tilde{S}%
_{t}^{T_{A},T_{M}}$ to have an irregular behaviour (jump to infinity) at
default of the entire pool. In terms of single name default times, the value
of the spread denominator is
\begin{equation*}
\Pi \left( \gamma _{t}^{T_{A},T_{M}}\right) =\sum_{i=1}^{n}\frac{1}{n}%
\sum_{j=A+1}^{M}\mathbb{E}\left[ \left. D\left( t,T_{j}\right) \alpha
_{j}1_{\left\{ \tau _{i}>T_{j}\right\} }\right| \mathcal{F}_{t}\right]
\end{equation*}
If we could price through a $\mathcal{H}_{t}^{k}$-expectation, analogously
to the single name case, we would exclude information on default of the $k$%
-th name. In this case $\mathbb{E}\left[ \left. D\left( t,T_{j}\right)
\alpha _{j}1_{\left\{ \tau _{k}>T_{j}\right\} }\right| \mathcal{H}_{t}^{k}%
\right] $ would never jump to zero, keeping the $\mathcal{H}_{t}^{k}$%
-expectation of the entire portfolio strictly positive. However, the choice
of a single pivotal name would be arbitrary, and, more importantly, using (%
\ref{formula_JeanRutk_defaultable}) requires that the payoff $Y_{t}^{T}$
goes to zero when $\tau _{k}\leq t$, which does not happen with the above
portfolio.

We find that the most effective possibility for using (\ref
{formula_JeanRutk_defaultable}) in this context is the following. We define
a sub-filtration that only excludes information on the event that sets $\Pi
\left( \gamma _{t}^{T_{A},T_{M}}\right) $ to zero. Define a new stopping
time
\begin{equation*}
\hat{\tau}=\max \left( \tau _{1},\tau _{2},\ldots ,\tau _{n}\right)
\end{equation*}
and define implicitly a new filtration $\mathcal{\hat{H}}_{t}$ such that
\begin{eqnarray*}
\mathcal{F}_{t} &=&\mathcal{\hat{J}}_{t}\vee \mathcal{\hat{H}}_{t} \\
\mathcal{\hat{J}}_{t} &=&\sigma \left( \left\{ \hat{\tau}>u\right\} ,u\leq
t\right) ,
\end{eqnarray*}
so that $\mathcal{\hat{H}}_{t}$ excludes, from the total flow of market
information, the information on the happening of a so-called portfolio
``armageddon event'', corresponding to the default of all names in a
portfolio. An extremely unlikely event when the option is on an a large
Index with high-rating names and the market situation is normal, but that
can have a non-negligible risk-neutral probability for products involving
optionality on smaller portfolios or in stressed market conditions, as we
show in Section \ref{Section_Tests}.

In the following we assume that
\begin{equation*}
\mathbb{Q}\left( \hat{\tau}>t|\mathcal{\hat{H}}_{t}\right) >0\text{ \ \ a.s.}
\end{equation*}
Exploiting that $\gamma _{t}^{T_{A},T_{M}}=\mathbf{1}_{\left\{ \hat{\tau}%
>T_{A}\right\} }\gamma _{t}^{T_{A},T_{M}}$, define
\begin{equation*}
\hat{\Pi}\left( \gamma _{t}^{T_{A},T_{M}}\right) :=\mathbb{E}\left[ \gamma
_{t}^{T_{A},T_{M}}|\mathcal{\hat{H}}_{t}\right]
\end{equation*}
so as to write
\begin{eqnarray}
\Pi \left( \gamma _{t}^{T_{A},T_{M}}\right) &=&\mathbb{E}\left[ \gamma
_{t}^{T_{A},T_{M}}\mathbf{|}\mathcal{F}_{t}\right] =\frac{\mathbf{1}%
_{\left\{ \hat{\tau}>t\right\} }}{\mathbb{Q}\left( \hat{\tau}>t|\mathcal{%
\hat{H}}_{t}\right) }\mathbb{E}\left[ \gamma _{t}^{T_{A},T_{M}}|\mathcal{%
\hat{H}}_{t}\right]  \label{formula_nonvanishing_numeraire} \\
&=&\frac{\mathbf{1}_{\left\{ \hat{\tau}>t\right\} }}{\mathbb{Q}\left( \hat{%
\tau}>t|\mathcal{\hat{H}}_{t}\right) }\hat{\Pi}\left( \gamma
_{t}^{T_{A},T_{M}}\right) .  \notag
\end{eqnarray}

The quantity $\hat{\Pi}\left( \gamma _{t}^{T_{A},T_{M}}\right) $ is never
null, and we will see that it is what we need for an effective definition of
the Index Spread and of an equivalent pricing measure for Index Options.

\subsection{Consistent computation of the Index Spread}

Now, for reaching a valid definition of the spread, there is one remaining
step. We need to price the Loss-Adjusted index (\ref
{formula_loss_adjusted_Index_payoff}) through the $\mathcal{\hat{H}}_{t}$
subfiltration. Unfortunately, we notice that applying (\ref
{formula_JeanRutk_defaultable}) with $\hat{\tau}$ playing the role of $\tau $
does not work, since (\ref{formula_definition_of_defaultable}) does not
hold. In fact the Loss-Adjusted index, differently from $\gamma
_{t}^{T_{A},T_{M}}$, is never null, because even in case $\hat{\tau}\leq
T_{A}$ we receive front end protection. Now we need a generalization of (\ref
{formula_JeanRutk_defaultable}). We can generalize (\ref
{formula_JeanRutk_defaultable}) as follows
\begin{eqnarray*}
\mathbf{Y}_{t}^{T} &=&\mathbf{1}_{\left\{ \hat{\tau}>t\right\} }\mathbf{Y}%
_{t}^{T}\mathbf{+1}_{\left\{ \hat{\tau}\leq t\right\} }\mathbf{Y}_{t}^{T}, \\
\mathbb{E}\left[ \mathbf{Y}_{t}^{T}\mathbf{|}\mathcal{F}_{t}\right] &=&%
\mathbb{E}\left[ \mathbf{1}_{\left\{ \hat{\tau}>t\right\} }\mathbf{Y}_{t}^{T}%
\mathbf{|}\mathcal{F}_{t}\right] +\mathbb{E}\left[ \mathbf{1}_{\left\{ \hat{%
\tau}\leq t\right\} }\mathbf{Y}_{t}^{T}\mathbf{|}\mathcal{F}_{t}\right]
\end{eqnarray*}
The first component can now be computed through (\ref
{formula_JeanRutk_defaultable}), since $\mathbf{1}_{\left\{ \hat{\tau}%
>t\right\} }\mathbf{Y}_{t}^{T}$ is a defaultable payoff:
\begin{equation*}
\mathbb{E}\left[ \mathbf{1}_{\left\{ \hat{\tau}>t\right\} }\mathbf{Y}_{t}^{T}%
\mathbf{|}\mathcal{F}_{t}\right] =\frac{\mathbf{1}_{\left\{ \hat{\tau}%
>t\right\} }}{\mathbb{Q}\left( \hat{\tau}>t|\mathcal{\hat{H}}_{t}\right) }%
\mathbb{E}\left[ \mathbf{1}_{\left\{ \hat{\tau}>t\right\} }\mathbf{Y}%
_{t}^{T}|\mathcal{\hat{H}}_{t}\right]
\end{equation*}
As for the second component,
\begin{eqnarray}
\mathbb{E}\left[ \mathbf{1}_{\left\{ \hat{\tau}\leq t\right\} }\mathbf{Y}%
_{t}^{T}\mathbf{|}\mathcal{F}_{t}\right] &=&\mathbb{E}\left[ \mathbf{1}%
_{\left\{ \hat{\tau}\leq t\right\} }\mathbf{Y}_{t}^{T}\mathbf{|}\mathcal{%
\hat{J}}_{t}\vee \mathcal{\hat{H}}_{t}\right] =\mathbb{E}\left[ \mathbf{1}%
_{\left\{ \hat{\tau}\leq t\right\} }\mathbf{Y}_{t}^{T}\mathbf{|}\sigma
\left( \hat{\tau}\right) \vee \mathcal{\hat{H}}_{t}\right]
\label{formula_JR_non_def_second_component} \\
&=&\mathbf{1}_{\left\{ \hat{\tau}\leq t\right\} }\mathbb{E}\left[ \mathbf{Y}%
_{t}^{T}\mathbf{|}\sigma \left( \hat{\tau}\right) \vee \mathcal{\hat{H}}_{t}%
\right]  \notag
\end{eqnarray}
\ See Bielecki and Rutkowski (2001) for more details on the above passages.
The final formula is
\begin{equation}
\mathbb{E}\left[ \mathbf{Y}_{t}^{T}\mathbf{|}\mathcal{F}_{t}\right] =\frac{%
\mathbf{1}_{\left\{ \hat{\tau}>t\right\} }}{\mathbb{Q}\left( \hat{\tau}>t|%
\mathcal{\hat{H}}_{t}\right) }\mathbb{E}\left[ \mathbf{1}_{\left\{ \hat{\tau}%
>t\right\} }\mathbf{Y}_{t}^{T}|\mathcal{\hat{H}}_{t}\right] +\mathbf{1}%
_{\left\{ \hat{\tau}\leq t\right\} }\mathbb{E}\left[ \mathbf{Y}_{t}^{T}%
\mathbf{|}\sigma \left( \hat{\tau}\right) \vee \mathcal{\hat{H}}_{t}\right] .
\label{formula_JeanRutk_non_defaultable2}
\end{equation}
The second component corresponds to the value of the payoff when we know
that all defaults have already happened, and we know the exact
``armageddon'' time. It is possible to use (\ref
{formula_JeanRutk_non_defaultable2}) for pricing a Loss-Adjusted index only
if we are able to compute the second component (\ref
{formula_JR_non_def_second_component}) without using assumptions on the
dynamics of the forward spread, that, analogously to the single name case,
we are going to give only for the $\mathcal{\hat{H}}_{t}$ filtration.

We now apply (\ref{formula_JeanRutk_non_defaultable2}) to the Loss-Adjusted
index payoff (\ref{formula_loss_adjusted_Index_payoff}):
\begin{eqnarray*}
&&\Pi \left( \tilde{I}_{t}^{T_{A},T_{M}}\left( K\right) \right) =\mathbf{1}%
_{\left\{ \hat{\tau}\leq t\right\} }\mathbb{E}\left[ \Phi
_{t}^{T_{A},T_{M}}-\Psi _{t}^{T_{A},T_{M}}\left( K\right) +F_{t}^{T_{A}}%
\mathbf{|}\sigma \left( \hat{\tau}\right) \vee \mathcal{\hat{H}}_{t}\right]
\\
&&+\frac{\mathbf{1}_{\left\{ \hat{\tau}>t\right\} }}{\mathbb{Q}\left( \hat{%
\tau}>t|\mathcal{\hat{H}}_{t}\right) }\mathbb{E}\left[ \mathbf{1}_{\left\{
\hat{\tau}>t\right\} }\Phi _{t}^{T_{A},T_{M}}-\mathbf{1}_{\left\{ \hat{\tau}%
>t\right\} }\Psi _{t}^{T_{A},T_{M}}\left( K\right) +\mathbf{1}_{\left\{ \hat{%
\tau}>t\right\} }F_{t}^{T_{A}}|\mathcal{\hat{H}}_{t}\right] \\
&=&:I_{1}+I_{2}
\end{eqnarray*}
We analyze first $I_{1}$. Notice that
\begin{equation*}
\mathbf{1}_{\left\{ \hat{\tau}\leq t\right\} }\mathbb{E}\left[ \Phi
_{t}^{T_{A},T_{M}}\mathbf{|}\sigma \left( \hat{\tau}\right) \vee \mathcal{%
\hat{H}}_{t}\right] =0,\;\;\;\;\;\;\;\mathbf{1}_{\left\{ \hat{\tau}\leq
t\right\} }\mathbb{E}\left[ \Psi _{t}^{T_{A},T_{M}}\left( K\right) \mathbf{|}%
\sigma \left( \hat{\tau}\right) \vee \mathcal{\hat{H}}_{t}\right] =0,
\end{equation*}
while
\begin{eqnarray*}
&&\mathbf{1}_{\left\{ \hat{\tau}\leq t\right\} }\mathbb{E}\left[ \left.
F_{t}^{T_{A}}\right| \mathbf{\sigma }\left( \hat{\tau}\right) \vee \mathcal{%
\hat{H}}_{t}\right] =\mathbf{1}_{\left\{ \hat{\tau}\leq t\right\} }\mathbb{E}%
\left[ \left. D\left( t,T_{A}\right) L\left( T_{A}\right) \right| \sigma
\left( \hat{\tau}\right) \vee \mathcal{\hat{H}}_{t}\right] \\
&=&\mathbf{1}_{\left\{ \hat{\tau}\leq t\right\} }\left( 1-R\right) \mathbb{E}%
\left[ \left. D\left( t,T_{A}\right) \right| \sigma \left( \hat{\tau}\right)
\vee \mathcal{\hat{H}}_{t}\right] =\mathbf{1}_{\left\{ \hat{\tau}\leq
t\right\} }\left( 1-R\right) P\left( t,T_{A}\right)
\end{eqnarray*}
Indeed, since we are conditioning to knowledge that all names have already
defaulted, the expectation of front end protection corresponds to the entire
notional diminished by the recovery.

We now analyze $I_{2}$. Notice that
\begin{eqnarray*}
\mathbf{1}_{\left\{ \hat{\tau}>t\right\} }\Phi _{t}^{T_{A},T_{M}} &=&\Phi
_{t}^{T_{A},T_{M}} \\
\mathbf{1}_{\left\{ \hat{\tau}>t\right\} }\Psi _{t}^{T_{A},T_{M}}\left(
K\right) &=&\Psi _{t}^{T_{A},T_{M}}\left( K\right)
\end{eqnarray*}
since protection and premium leg payoffs go to zero when all names in the
portfolio default, so, calling
\begin{equation*}
\mathbb{E}\left[ \Phi _{t}^{T_{A},T_{M}}|\mathcal{\hat{H}}_{t}\right] =:\hat{%
\Pi}\left( \Phi _{t}^{T_{A},T_{M}}\right) ,\;\;\;\;\;\mathbb{E}\left[ \Psi
_{t}^{T_{A},T_{M}}\left( K\right) |\mathcal{\hat{H}}_{t}\right] =:\hat{\Pi}%
\left( \Psi _{t}^{T_{A},T_{M}}\left( K\right) \right)
\end{equation*}
we have
\begin{eqnarray}
&&\Pi \left( \tilde{I}_{t}^{T_{A},T_{M}}\left( K\right) \right) =\mathbf{1}%
_{\left\{ \hat{\tau}\leq t\right\} }\left( 1-R\right) P\left( t,T_{A}\right)
\label{formula_index_Hprice} \\
&&+\frac{\mathbf{1}_{\left\{ \hat{\tau}>t\right\} }}{\mathbb{Q}\left( \hat{%
\tau}>t|\mathcal{\hat{H}}_{t}\right) }\left\{ \hat{\Pi}\left( \Phi
_{t}^{T_{A},T_{M}}\right) -\hat{\Pi}\left( \Psi _{t}^{T_{A},T_{M}}\left(
K\right) \right) +\mathbb{E}\left[ \mathbf{1}_{\left\{ \hat{\tau}>t\right\}
}F_{t}^{T_{A}}|\mathcal{\hat{H}}_{t}\right] \right\}  \notag
\end{eqnarray}
Now it is convenient to perform the following decomposition:
\begin{eqnarray*}
\mathbb{E}\left[ \mathbf{1}_{\left\{ \hat{\tau}>t\right\} }F_{t}^{T_{A}}|%
\mathcal{\hat{H}}_{t}\right] &=&\mathbb{E}\left[ \mathbf{1}_{\left\{ t<\hat{%
\tau}\leq T_{A}\right\} }F_{t}^{T_{A}}|\mathcal{\hat{H}}_{t}\right] +\mathbb{%
E}\left[ \mathbf{1}_{\left\{ \hat{\tau}>T_{A}\right\} }F_{t}^{T_{A}}|%
\mathcal{\hat{H}}_{t}\right] \\
&=&\left( 1-R\right) \mathbb{E}\left[ \mathbf{1}_{\left\{ t<\hat{\tau}\leq
T_{A}\right\} }D\left( t,T_{A}\right) |\mathcal{\hat{H}}_{t}\right] +\mathbb{%
E}\left[ \mathbf{1}_{\left\{ \hat{\tau}>T_{A}\right\} }F_{t}^{T_{A}}|%
\mathcal{\hat{H}}_{t}\right]
\end{eqnarray*}
so that
\begin{eqnarray}
&&\Pi \left( \tilde{I}_{t}^{T_{A},T_{M}}\left( K\right) \right)
\begin{tabular}{l}
$=$%
\end{tabular}
\label{formula_index_Hprice_decomposed} \\
&=&\frac{\mathbf{1}_{\left\{ \hat{\tau}>t\right\} }}{\mathbb{Q}\left( \hat{%
\tau}>t|\mathcal{\hat{H}}_{t}\right) }\left\{ \hat{\Pi}\left( \Phi
_{t}^{T_{A},T_{M}}\right) -\hat{\Pi}\left( \Psi _{t}^{T_{A},T_{M}}\left(
K\right) \right) +\mathbb{E}\left[ \mathbf{1}_{\left\{ \hat{\tau}%
>T_{A}\right\} }F_{t}^{T_{A}}|\mathcal{\hat{H}}_{t}\right] \right\}  \notag
\\
&&+\frac{\mathbf{1}_{\left\{ \hat{\tau}>t\right\} }\left( 1-R\right) }{%
\mathbb{Q}\left( \hat{\tau}>t|\mathcal{\hat{H}}_{t}\right) }\mathbb{\times }%
\mathbb{E}\left[ \mathbf{1}_{\left\{ t<\hat{\tau}\leq T_{A}\right\} }D\left(
t,T_{A}\right) |\mathcal{\hat{H}}_{t}\right] +  \notag \\
&&+\mathbf{1}_{\left\{ \hat{\tau}\leq t\right\} }\left( 1-R\right) P\left(
t,T_{A}\right)  \notag
\end{eqnarray}
Formula (\ref{formula_index_Hprice_decomposed}) shows the actual components
of the value of the Loss-Adjusted index, and will lead us in the next
passages required for reaching a consistent valuation of the Index Option.
We see that in a Loss-Adjusted portfolio we cannot define in all scenarios
the equilibrium spread as the value of the spread setting the index value (%
\ref{formula_index_Hprice}) to zero, since when $\left\{ \hat{\tau}\leq
t\right\} $ the index value will always be $\left( 1-R\right) P\left(
t,T_{A}\right) $. Therefore a consistent definition of the Index spread must
avoid the hopeless attempt of globally setting to zero a value that in some
scenarios can never be zero.

The financially meaningful definition of the Index Spread, that, as we will
see in Section \ref{section_equivalent_measure}, also makes it a martingale
under a natural pricing measure, considers the level of $K$ setting the
Index value to zero in all scenarios where some names survive until
maturity. Only in such scenarios, in fact, the payoff actually depends on
the Index Spread. This corresponds to setting to zero only the first
component of the index value (\ref{formula_index_Hprice_decomposed}),
\begin{equation*}
\frac{\mathbf{1}_{\left\{ \hat{\tau}>t\right\} }}{\mathbb{Q}\left( \hat{\tau}%
>t|\mathcal{\hat{H}}_{t}\right) }\left\{ \hat{\Pi}\left( \Phi
_{t}^{T_{A},T_{M}}\right) -\hat{\Pi}\left( \Psi _{t}^{T_{A},T_{M}}\left(
K\right) \right) +\mathbb{E}\left[ \mathbf{1}_{\left\{ \hat{\tau}%
>T_{A}\right\} }F_{t}^{T_{A}}|\mathcal{\hat{H}}_{t}\right] \right\} ,
\end{equation*}
which is the price of an armageddon-knock out tradable asset. We obtain the
following definition of the equilibrium \emph{Arbitrage-free Index Spread}
\begin{equation}
\;\ \ \ \ \ \ \ \ \ \ \ \ \ \ \ \ \ \ \ \;\;\;\hat{S}_{t}^{T_{A},T_{M}}=%
\frac{\hat{\Pi}\left( \Phi _{t}^{T_{A},T_{M}}\right) +\mathbb{E}\left[
\mathbf{1}_{\left\{ \hat{\tau}>T_{A}\right\} }F_{t}^{T_{A}}|\mathcal{\hat{H}}%
_{t}\right] }{\hat{\Pi}\left( \gamma _{t}^{T_{A},T_{M}}\right) }
\label{formula_right_spreaddefinition}
\end{equation}
This definition of the index spread is both regular, since $\hat{\Pi}\left(
\gamma _{t}^{T_{A},T_{M}}\right) $ is bounded away from zero, and its
definition has a reasonable financial meaning, since it does not claim to
set to zero a value that cannot be set to zero with probability 1.

\section{Arbitrage-free pricing of Index Options\label{Section_IndexOption}}

The Index Option has payoff
\begin{equation*}
Y_{t}^{T_{A},T_{M}}\left( K\right) =D\left( t,T_{A}\right) \left( \Pi \left(
\tilde{I}_{T_{A}}^{T_{A},T_{M}}\left( K\right) \right) \right) ^{+}.
\end{equation*}
In this section we evaluate the option, through change to a newly defined
pricing measure. However we first need to analyze in more details the payoff.

\subsection{Analysis of the Index Option Payoff}

From (\ref{formula_index_Hprice}) and (\ref{formula_right_spreaddefinition}%
),
\begin{equation*}
\Pi \left( \tilde{I}_{T_{A}}^{T_{A},T_{M}}\left( K\right) \right) =\frac{%
\mathbf{1}_{\left\{ \hat{\tau}>T_{A}\right\} }}{\mathbb{Q}\left( \hat{\tau}%
>T_{A}|\mathcal{\hat{H}}_{T_{A}}\right) }\hat{\Pi}\left( \gamma
_{T_{A}}^{T_{A},T_{M}}\right) \left( \hat{S}_{T_{A}}^{T_{A},T_{M}}-K\right) +%
\mathbf{1}_{\left\{ \hat{\tau}\leq T_{A}\right\} }\left( 1-R\right)
\end{equation*}
The use of subfiltrations and pricing formula (\ref
{formula_JeanRutk_non_defaultable2}) induces a redefinition not only of the
underlying spread, but also of the value of the Index at Maturity and thus
of the Index Option payoff:
\begin{eqnarray*}
Y_{t}^{T_{A},T_{M}}\left( K\right) &=&D\left( t,T_{A}\right) \left( \frac{%
\mathbf{1}_{\left\{ \hat{\tau}>T_{A}\right\} }}{\mathbb{Q}\left( \hat{\tau}%
>T_{A}|\mathcal{\hat{H}}_{T_{A}}\right) }\hat{\Pi}\left( \gamma
_{T_{A}}^{T_{A},T_{M}}\right) \left( \hat{S}_{T_{A}}^{T_{A},T_{M}}-K\right) +%
\mathbf{1}_{\left\{ \hat{\tau}\leq T_{A}\right\} }\left( 1-R\right) \right)
^{+} \\
&=&D\left( t,T_{A}\right) \frac{\mathbf{1}_{\left\{ \hat{\tau}>T_{A}\right\}
}}{\mathbb{Q}\left( \hat{\tau}>T_{A}|\mathcal{\hat{H}}_{T_{A}}\right) }\hat{%
\Pi}\left( \gamma _{T_{A}}^{T_{A},T_{M}}\right) \left( \hat{S}%
_{T_{A}}^{T_{A},T_{M}}-K\right) ^{+}+\mathbf{1}_{\left\{ \hat{\tau}\leq
T_{A}\right\} }\left( 1-R\right) ,
\end{eqnarray*}
where the last passage follows from the properties of indicators. Notice the
correct payoff is split in two parts, and differently from (\ref
{formula_wrong3_option_payoff}) it does not lead to Problem 2 of Section \ref
{Section_advantages&problems}, but instead it represents the actual payoff
in all states of the world. Now, using again pricing formula (\ref
{formula_JeanRutk_non_defaultable2}), we can reach a fully consistent
computation of the value of the Credit Index Option.
\begin{eqnarray*}
\Pi \left( Y_{t}^{T_{A},T_{M}}\left( K\right) \right) &=&\mathbf{1}_{\left\{
\hat{\tau}\leq t\right\} }\mathbb{E}\left[ Y_{t}^{T_{A},T_{M}}\left(
K\right) |\sigma \left( \hat{\tau}\right) \vee \mathcal{\hat{H}}_{t}\right]
\\
&&+\frac{\mathbf{1}_{\left\{ \hat{\tau}>t\right\} }}{\mathbb{Q}\left( \hat{%
\tau}>t|\mathcal{\hat{H}}_{t}\right) }\mathbb{E}\left[ \mathbf{1}_{\left\{
\hat{\tau}>t\right\} }Y_{t}^{T_{A},T_{M}}\left( K\right) |\mathcal{\hat{H}}%
_{t}\right] \\
&=&O_{1}+O_{2}
\end{eqnarray*}
Notice $\mathbf{1}_{\left\{ \hat{\tau}\leq t\right\} }\mathbb{\times }%
\mathbf{1}_{\left\{ \hat{\tau}>T_{A}\right\} }=0$, $\mathbf{1}_{\left\{ \hat{%
\tau}\leq t\right\} }\mathbb{\times }\mathbf{1}_{\left\{ \hat{\tau}\leq
T_{A}\right\} }=\mathbf{1}_{\left\{ \hat{\tau}\leq t\right\} }$, so
\begin{eqnarray*}
O_{1} &=&\mathbf{1}_{\left\{ \hat{\tau}\leq t\right\} }\mathbf{1}_{\left\{
\hat{\tau}\leq T_{A}\right\} }\mathbb{E}\left[ D\left( t,T_{A}\right) \left(
1-R\right) |\sigma \left( \hat{\tau}\right) \vee \mathcal{\hat{H}}_{t}\right]
\\
&=&\mathbf{1}_{\left\{ \hat{\tau}\leq t\right\} }\left( 1-R\right) P\left(
t,T_{A}\right) \text{,}
\end{eqnarray*}
and
\begin{eqnarray*}
&&O_{2}=\frac{\mathbf{1}_{\left\{ \hat{\tau}>t\right\} }}{\mathbb{Q}\left(
\hat{\tau}>t|\mathcal{\hat{H}}_{t}\right) }\mathbb{E}\left[ D\left(
t,T_{A}\right) \frac{\mathbf{1}_{\left\{ \hat{\tau}>T_{A}\right\} }}{\mathbb{%
Q}\left( \hat{\tau}>T_{A}|\mathcal{\hat{H}}_{T_{A}}\right) }\hat{\Pi}\left(
\gamma _{T_{A}}^{T_{A},T_{M}}\right) \left( \hat{S}_{T_{A}}^{T_{A},T_{M}}-K%
\right) ^{+}|\mathcal{\hat{H}}_{t}\right] + \\
&&\frac{\mathbf{1}_{\left\{ \hat{\tau}>t\right\} }}{\mathbb{Q}\left( \hat{%
\tau}>t|\mathcal{\hat{H}}_{t}\right) }\mathbb{E}\left[ D\left(
t,T_{A}\right) \mathbf{1}_{\left\{ t<\hat{\tau}\leq T_{A}\right\} }\left(
1-R\right) |\mathcal{\hat{H}}_{t}\right]
\end{eqnarray*}
This leads to the option pricing formula
\begin{eqnarray}
&&\Pi \left( Y_{t}^{T_{A},T_{M}}\left( K\right) \right)
\label{formula_option_price} \\
&=&\frac{\mathbf{1}_{\left\{ \hat{\tau}>t\right\} }}{\mathbb{Q}\left( \hat{%
\tau}>t|\mathcal{\hat{H}}_{t}\right) }\mathbb{E}\left[ D\left(
t,T_{A}\right) \frac{\mathbf{1}_{\left\{ \hat{\tau}>T_{A}\right\} }}{\mathbb{%
Q}\left( \hat{\tau}>T_{A}|\mathcal{\hat{H}}_{T_{A}}\right) }\hat{\Pi}\left(
\gamma _{T_{A}}^{T_{A},T_{M}}\right) \left( \hat{S}_{T_{A}}^{T_{A},T_{M}}-K%
\right) ^{+}|\mathcal{\hat{H}}_{t}\right]  \notag \\
&&+\frac{\mathbf{1}_{\left\{ \hat{\tau}>t\right\} }}{\mathbb{Q}\left( \hat{%
\tau}>t|\mathcal{\hat{H}}_{t}\right) }\mathbb{E}\left[ D\left(
t,T_{A}\right) \mathbf{1}_{\left\{ t<\hat{\tau}\leq T_{A}\right\} }\left(
1-R\right) |\mathcal{\hat{H}}_{t}\right]  \notag \\
&&+\mathbf{1}_{\left\{ \hat{\tau}\leq t\right\} }\left( 1-R\right) P\left(
t,T_{A}\right)  \notag \\
&=&:\hat{O}_{1}+\hat{O}_{2}+\hat{O}_{3}  \notag
\end{eqnarray}
So we have reached a formula that really shows the different components of
the option value, and allows us to compute them in a convenient way. The
third part $\hat{O}_{3}$ is just the present value of the option when a
portfolio ``armageddon event'' happens before $t$. The second part $\hat{O}%
_{2}$ takes correctly into account the probability of such an event between
now and the option expiry. The first part $\hat{O}_{1}$, that erroneously
was the only one considered in the simpler formula (\ref
{formula_wrong2_option_price}), is the part that we will compute through a
closed-form standard option formula.

This requires the definition of a viable change of measure, therefore it
means solving also Problem 3. We see in the next section that, although this
is technically the most demanding of the three problems, the preceding
analysis and in particular the introduction of an appropriate subfiltration
already gives us the correct tools to deal with this issue.

\subsection{An Equivalent measure for Credit Multiname Options \label%
{section_equivalent_measure}\label{Section_equivalentmeasure}}

Now we deal with the evaluation of the first part of (\ref
{formula_option_price}), that as we have seen is the only part that contains
optionality. First of all we compute
\begin{eqnarray*}
\hat{O}_{1} &=&\frac{\mathbf{1}_{\left\{ \hat{\tau}>t\right\} }}{\mathbb{Q}%
\left( \hat{\tau}>t|\mathcal{\hat{H}}_{t}\right) }\mathbb{E}\left[ D\left(
t,T_{A}\right) \frac{\mathbf{1}_{\left\{ \hat{\tau}>T_{A}\right\} }}{\mathbb{%
Q}\left( \hat{\tau}>T_{A}|\mathcal{\hat{H}}_{T_{A}}\right) }\hat{\Pi}\left(
\gamma _{T_{A}}^{T_{A},T_{M}}\right) \left( \hat{S}_{T_{A}}^{T_{A},T_{M}}-K%
\right) ^{+}|\mathcal{\hat{H}}_{t}\right] \\
&=&\frac{\mathbf{1}_{\left\{ \hat{\tau}>t\right\} }}{\mathbb{Q}\left( \hat{%
\tau}>t|\mathcal{\hat{H}}_{t}\right) }\mathbb{E}\left[ \mathbb{E}\left[
D\left( t,T_{A}\right) \frac{\mathbf{1}_{\left\{ \hat{\tau}>T_{A}\right\} }}{%
\mathbb{Q}\left( \hat{\tau}>T_{A}|\mathcal{\hat{H}}_{T_{A}}\right) }\hat{\Pi}%
\left( \gamma _{T_{A}}^{T_{A},T_{M}}\right) \left( \hat{S}%
_{T_{A}}^{T_{A},T_{M}}-K\right) ^{+}|\mathcal{\hat{H}}_{T_{A}}\right] |%
\mathcal{\hat{H}}_{t}\right] \\
&=&\frac{\mathbf{1}_{\left\{ \hat{\tau}>t\right\} }}{\mathbb{Q}\left( \hat{%
\tau}>t|\mathcal{\hat{H}}_{t}\right) }\mathbb{E}\left[ D\left(
t,T_{A}\right) \frac{1}{\mathbb{Q}\left( \hat{\tau}>T_{A}|\mathcal{\hat{H}}%
_{T_{A}}\right) }\hat{\Pi}\left( \gamma _{T_{A}}^{T_{A},T_{M}}\right) \left(
\hat{S}_{T_{A}}^{T_{A},T_{M}}-K\right) ^{+}\mathbb{E}\left[ \mathbf{1}%
_{\left\{ \hat{\tau}>T_{A}\right\} }|\mathcal{\hat{H}}_{T_{A}}\right] |%
\mathcal{\hat{H}}_{t}\right]
\end{eqnarray*}
Since
\begin{equation*}
\mathbb{E}\left[ \mathbf{1}_{\left\{ \hat{\tau}>T_{A}\right\} }|\mathcal{%
\hat{H}}_{T_{A}}\right] =\mathbb{Q}\left( \hat{\tau}>T_{A}|\mathcal{\hat{H}}%
_{T_{A}}\right)
\end{equation*}
we have
\begin{equation*}
\hat{O}_{1}=\frac{\mathbf{1}_{\left\{ \hat{\tau}>t\right\} }}{\mathbb{Q}%
\left( \hat{\tau}>t|\mathcal{\hat{H}}_{t}\right) }\mathbb{E}\left[ D\left(
t,T_{A}\right) \hat{\Pi}\left( \gamma _{T_{A}}^{T_{A},T_{M}}\right) \left(
\hat{S}_{T_{A}}^{T_{A},T_{M}}-K\right) ^{+}|\mathcal{\hat{H}}_{t}\right]
\end{equation*}
Now it is natural to take the quantity
\begin{equation*}
\hat{\Pi}\left( \gamma _{t}^{T_{A},T_{M}}\right) =\mathbb{E}\left[ \gamma
_{t}^{T_{A},T_{M}}|\mathcal{\hat{H}}_{t}\right] =\mathbb{E}\left[
\sum_{j=A+1}^{M}D\left( t,T_{j}\right) \alpha _{j}\left( 1-\frac{L\left(
T_{j}\right) }{\left( 1-R\right) }\right) |\mathcal{\hat{H}}_{t}\right]
\end{equation*}
to define a probability measure $\mathbb{\hat{Q}}^{T_{A},T_{M}}$ allowing to
simplify the computation. In fact, differently from $\Pi \left( \gamma
_{t}^{T_{A},T_{M}}\right) $ which would be the quantity that one should
select if subfiltrations had not been introduced, $\hat{\Pi}\left( \gamma
_{t}^{T_{A},T_{M}}\right) $ is strictly positive. Only with a strictly
positive quantity we remain in the context of probability measures which are
equivalent to the risk neutral measure and, by transitivity, to the real
world probability measure.

We define the $T_{A},T_{M}$\emph{-no-armageddon pricing measure} $\mathbb{%
\hat{Q}}^{T_{A},T_{M}}$ through definition of the Radon-Nykodim derivative
of this measure with respect to $\mathbb{Q}$ , restricted to $\mathcal{\hat{H%
}}_{T_{A}}$,
\begin{equation}
Z_{T_{A}}=\left. \frac{d\mathbb{\hat{Q}}^{T_{A},T_{M}}}{d\mathbb{Q}}\right|
_{\mathcal{\hat{H}}_{T_{A}}}.  \label{formula_restricted_radon_nykodim}
\end{equation}
Recall that, by definition of Radon-Nykodim derivative,
\begin{equation*}
\mathbb{\hat{Q}}^{T_{A},T_{M}}\left( A\right) =\int_{A}Z_{T_{A}}d\mathbb{Q}%
,\;\;\;A\in \mathcal{\hat{H}}_{T_{A}}.
\end{equation*}
The definition of $Z_{T_{A}}$ that allows a computational simplification of
option pricing is
\begin{equation*}
Z_{T_{A}}=\frac{B_{0}\;\hat{\Pi}\left( \gamma _{T_{A}}^{T_{A},T_{M}}\right)
}{\hat{\Pi}\left( \gamma _{0}^{T_{A},T_{M}}\right) \;B_{T_{A}}}.
\end{equation*}
Notice that $B_{t}$, the locally risk-free bank account, is $\mathcal{\hat{H}%
}_{t}$-adapted. For derivatives pricing we need to define the restriction of
$\mathbb{\hat{Q}}^{T_{A},T_{M}}$ to $\mathcal{\hat{H}}_{t}$, $t\leq T_{A}$.
It is clear that if we define $Z_{t}$, $t\leq T_{A}$, as an $\mathcal{\hat{H}%
}_{t}$ martingale
\begin{equation*}
Z_{t}=\mathbb{E}\left[ \left. \left. \frac{d\mathbb{\hat{Q}}^{T_{A},T_{M}}}{d%
\mathbb{Q}}\right| _{\mathcal{\hat{H}}_{T_{A}}}\right| \mathcal{\hat{H}}_{t}%
\right]
\end{equation*}
we have that $Z_{t}$ is exactly the restriction we are looking for, $%
Z_{t}=\left. \frac{d\mathbb{\hat{Q}}^{T_{A},T_{M}}}{d\mathbb{Q}}\right| _{%
\mathcal{\hat{H}}_{t}}$, since for all $C\in \mathcal{\hat{H}}_{t}$
\begin{equation*}
\int_{C}\left. \frac{d\mathbb{\hat{Q}}^{T_{A},T_{M}}}{d\mathbb{Q}}\right| _{%
\mathcal{\hat{H}}_{t}}d\mathbb{Q}=\int_{C}\mathbb{E}\left[ \left. \left.
\frac{d\mathbb{\hat{Q}}^{T_{A},T_{M}}}{d\mathbb{Q}}\right| _{\mathcal{\hat{H}%
}_{T_{A}}}\right| \mathcal{\hat{H}}_{t}\right] d\mathbb{Q}=\int_{C}\left.
\frac{d\mathbb{\hat{Q}}^{T_{A},T_{M}}}{d\mathbb{Q}}\right| _{\mathcal{\hat{H}%
}_{T_{A}}}d\mathbb{Q}=\mathbb{\hat{Q}}^{T_{A},T_{M}}\left( C\right) \text{.}
\end{equation*}
where the last but one passage comes from Kolmogorov's definition of
conditional expectation. Then notice that
\begin{eqnarray*}
\left. \frac{d\mathbb{\hat{Q}}^{T_{A},T_{M}}}{d\mathbb{Q}}\right| _{\mathcal{%
\hat{H}}_{t}} &=&\mathbb{E}\left[ \left. \frac{B_{0}\;\hat{\Pi}\left( \gamma
_{T_{A}}^{T_{A},T_{M}}\right) }{\hat{\Pi}\left( \gamma
_{0}^{T_{A},T_{M}}\right) \;B_{T_{A}}}\right| \mathcal{\hat{H}}_{t}\right] \\
&=&\frac{B_{0}}{\hat{\Pi}\left( \gamma _{0}^{T_{A},T_{M}}\right) }\mathbb{E}%
\left[ \left. \mathbb{E}\left[ \sum_{j=A+1}^{M}\frac{1}{B_{T_{j}}}\alpha
_{j}\left( 1-\frac{L\left( T_{j}\right) }{\left( 1-R\right) }\right) |%
\mathcal{\hat{H}}_{T_{A}}\right] \right| \mathcal{\hat{H}}_{t}\right] \\
&=&\frac{B_{0}}{\hat{\Pi}\left( \gamma _{0}^{T_{A},T_{M}}\right) B_{t}}%
\mathbb{E}\left[ \left. \sum_{j=A+1}^{M}\frac{B_{t}}{B_{T_{j}}}\alpha
_{j}\left( 1-\frac{L\left( T_{j}\right) }{\left( 1-R\right) }\right) \right|
\mathcal{\hat{H}}_{t}\right] \\
&=&\frac{B_{0}\hat{\Pi}\left( \gamma _{t}^{T_{A},T_{M}}\right) }{\hat{\Pi}%
\left( \gamma _{0}^{T_{A},T_{M}}\right) B_{t}}
\end{eqnarray*}
so also the Radon-Nykodim derivative restricted to all $\mathcal{\hat{H}}%
_{t} $, $t\leq T_{A}$, can be expressed in closed form through market
quantities. This is sufficient to apply the well-known Bayes rule for
conditional change of measure.

Consider a sub $\sigma $-algebra $\mathcal{N}$ of $\sigma $-algebra $%
\mathcal{M}$ and an $\mathcal{M}$-measurable $X$, integrable under the
measures $P1$ and $P2$, $P1\sim P2$. We have the following result.

\begin{proposition}
\label{Theorem}(Bayes rule for conditional change of measure) When $X$ is $%
\mathcal{M}$-measurable
\begin{equation*}
\mathbb{E}^{P2}\left[ X|\mathcal{N}\right] =\mathbb{E}^{P1}\left[ \left. X%
\frac{\mathbb{E}^{P1}\left[ \frac{dP2}{dP1}|\mathcal{M}\right] }{\mathbb{E}%
^{P1}\left[ \frac{dP2}{dP1}|\mathcal{N}\right] }\right| \mathcal{N}\right] .
\end{equation*}
\end{proposition}

\begin{proof}
\noindent The RHS is by definition $\mathcal{N}$-measurable. Exploiting
\begin{equation*}
\mathbb{E}^{P2}\left[ 1_{C}X\right] =\mathbb{E}^{P1}\left[ 1_{C}X\mathbb{E}%
^{P1}\left[ \frac{dP2}{dP1}|\mathcal{N}\right] \right] ,
\end{equation*}
for any $C\in \mathcal{N}$%
\begin{eqnarray*}
&&\int_{C}\mathbb{E}^{P1}\left[ \left. X\mathbb{E}^{P1}\left[ \frac{dP2}{dP1}%
|\mathcal{M}\right] \frac{1}{\mathbb{E}^{P1}\left[ \frac{dP2}{dP1}|\mathcal{N%
}\right] }\right| \mathcal{N}\right] dP2 \\
&=&\int_{C}\mathbb{E}^{P1}\left[ \left. X\mathbb{E}^{P1}\left[ \frac{dP2}{dP1%
}|\mathcal{M}\right] \right| \mathcal{N}\right] dP1=\int_{C}\mathbb{E}^{P1}%
\left[ \left. X\frac{dP2}{dP1}\right| \mathcal{N}\right] dP1
\end{eqnarray*}
By definition of conditional expectation,
\begin{equation*}
=\int_{C}\mathbb{E}^{P1}\left[ \left. X\frac{dP2}{dP1}\right| \mathcal{N}%
\right] dP1=\int_{C}X\frac{dP2}{dP1}dP1=\int_{C}XdP2.
\end{equation*}
\end{proof}

In our context, we have to compute
\begin{equation*}
\hat{O}_{1}=\frac{\mathbf{1}_{\left\{ \hat{\tau}>t\right\} }}{\mathbb{Q}%
\left( \hat{\tau}>t|\mathcal{\hat{H}}_{t}\right) }\mathbb{E}\left[ D\left(
t,T_{A}\right) \hat{\Pi}\left( \gamma _{T_{A}}^{T_{A},T_{M}}\right) \left(
\hat{S}_{T_{A}}^{T_{A},T_{M}}-K\right) ^{+}|\mathcal{\hat{H}}_{t}\right] .
\end{equation*}
and we have
\begin{equation*}
\frac{\mathbb{E}^{P1}\left[ \frac{dP2}{dP1}|\mathcal{M}\right] }{\mathbb{E}%
^{P1}\left[ \frac{dP2}{dP1}|\mathcal{N}\right] }=\frac{\mathbb{E}\left[
\frac{d\mathbb{\hat{Q}}^{T_{A},T_{M}}}{d\mathbb{Q}}|\mathcal{\hat{H}}_{T_{A}}%
\right] }{\mathbb{E}\left[ \frac{d\mathbb{\hat{Q}}^{T_{A},T_{M}}}{d\mathbb{Q}%
}|\mathcal{\hat{H}}_{t}\right] }=\frac{B_{t}\;\hat{\Pi}\left( \gamma
_{T_{A}}^{T_{A},T_{M}}\right) }{B_{T_{A}}\;\hat{\Pi}\left( \gamma
_{t}^{T_{A},T_{M}}\right) }.
\end{equation*}
Thus
\begin{equation*}
\hat{O}_{1}=\frac{\mathbf{1}_{\left\{ \hat{\tau}>t\right\} }}{\mathbb{Q}%
\left( \hat{\tau}>t|\mathcal{\hat{H}}_{t}\right) }\hat{\Pi}\left( \gamma
_{t}^{T_{A},T_{M}}\right) \mathbb{E}\left[ \frac{\mathbb{E}\left[ \frac{d%
\mathbb{\hat{Q}}^{T_{A},T_{M}}}{d\mathbb{Q}}|\mathcal{\hat{H}}_{T_{A}}\right]
}{\mathbb{E}\left[ \frac{d\mathbb{\hat{Q}}^{T_{A},T_{M}}}{d\mathbb{Q}}|%
\mathcal{\hat{H}}_{t}\right] }\left( \hat{S}_{T_{A}}^{T_{A},T_{M}}-K\right)
^{+}|\mathcal{\hat{H}}_{t}\right] .
\end{equation*}
Noticing that $\left( \hat{S}_{T_{A}}^{T_{A},T_{M}}-K\right) ^{+}$ is $%
\mathcal{\hat{H}}_{T_{A}}$-measurable, we apply Proposition \ref{Theorem},
\begin{eqnarray}
\hat{O}_{1} &=&\frac{\mathbf{1}_{\left\{ \hat{\tau}>t\right\} }}{\mathbb{Q}%
\left( \hat{\tau}>t|\mathcal{\hat{H}}_{t}\right) }\hat{\Pi}\left( \gamma
_{t}^{T_{A},T_{M}}\right) \mathbb{\hat{E}}^{T_{A},T_{M}}\left[ \left( \hat{S}%
_{T_{A}}^{T_{A},T_{M}}-K\right) ^{+}|\mathcal{\hat{H}}_{t}\right]
\label{formula_optionpricing_beforelognormality} \\
&=&\Pi \left( \gamma _{t}^{T_{A},T_{M}}\right) \mathbb{\hat{E}}^{T_{A},T_{M}}%
\left[ \left( \hat{S}_{T_{A}}^{T_{A},T_{M}}-K\right) ^{+}|\mathcal{\hat{H}}%
_{t}\right]   \notag
\end{eqnarray}
Now we analyze the behaviour of the underlying spread under $\mathbb{\hat{Q}}%
^{T_{A},T_{M}}$%
\begin{eqnarray*}
&&\mathbb{\hat{E}}^{T_{A},T_{M}}\left[ \hat{S}_{T_{A}}^{T_{A},T_{M}}|%
\mathcal{\hat{H}}_{t}\right] =\mathbb{E}\left[ \left. \hat{S}%
_{T_{A}}^{T_{A},T_{M}}\frac{B_{t}\;\hat{\Pi}\left( \gamma
_{T_{A}}^{T_{A},T_{M}}\right) }{B_{T_{A}}\;\hat{\Pi}\left( \gamma
_{t}^{T_{A},T_{M}}\right) }\right| \mathcal{\hat{H}}_{t}\right]  \\
&=&\frac{\mathbb{E}\left[ \left. \mathbb{E}\left[ \Phi _{t}^{T_{A},T_{M}}+%
\mathbf{1}_{\left\{ \hat{\tau}>T_{A}\right\} }D\left( t,T_{A}\right) L\left(
T_{A}\right) |\mathcal{\hat{H}}_{T_{A}}\right] \right| \mathcal{\hat{H}}_{t}%
\right] }{\hat{\Pi}\left( \gamma _{t}^{T_{A},T_{M}}\right) } \\
&=&\frac{\mathbb{E}\left[ \Phi _{t}^{T_{A},T_{M}}+\mathbf{1}_{\left\{ \hat{%
\tau}>T_{A}\right\} }D\left( t,T_{A}\right) L\left( T_{A}\right) |\mathcal{%
\hat{H}}_{t}\right] }{\hat{\Pi}\left( \gamma _{t}^{T_{A},T_{M}}\right) } \\
&=&\frac{\hat{\Pi}\left( \Phi _{t}^{T_{A},T_{M}}\right) +\mathbb{E}\left[
\mathbf{1}_{\left\{ \hat{\tau}>T_{A}\right\} }F_{t}^{T_{A}}|\mathcal{\hat{H}}%
_{t}\right] }{\hat{\Pi}\left( \gamma _{t}^{T_{A},T_{M}}\right) }=\hat{S}%
_{t}^{T_{A},T_{M}}
\end{eqnarray*}
So $\hat{S}_{t}^{T_{A},T_{M}}$ is a $\mathcal{\hat{H}}_{t}$-martingale under
$\mathbb{\hat{Q}}^{T_{A},T_{M}}$. If we assume
\begin{equation*}
d\hat{S}_{t}^{T_{A},T_{M}}=\hat{\sigma}^{T_{A},T_{M}}\hat{S}%
_{t}^{T_{A},T_{M}}dV^{T_{A},T_{M}},\;\;t\leq T_{a}
\end{equation*}
where $V^{T_{A},T_{M}}$ is a standard brownian motion under $\mathbb{\hat{Q}}%
^{T_{A},T_{M}}$ and $\hat{\sigma}^{T_{A},T_{M}}$ is the instantaneous
volatility, we have the following \emph{Arbitrage-free Credit Index Option
formula}
\begin{equation}
\;\;\;\;\;\;\;\;\;\;\;\;\;\;\Pi \left( Y_{t}^{T_{A},T_{M}}\left( K\right)
\right) =\Pi \left( \gamma _{t}^{T_{A},T_{M}}\right) Black\left( \hat{S}%
_{t}^{T_{A},T_{M}},K,\hat{\sigma}^{T_{A},T_{M}}\sqrt{T_{A}-t}\right)
\vspace*{-0.37cm}  \label{formula_right_option_price}
\end{equation}
\begin{eqnarray*}
&&+\frac{\mathbf{1}_{\left\{ \hat{\tau}>t\right\} }}{\mathbb{Q}\left( \hat{%
\tau}>t|\mathcal{\hat{H}}_{t}\right) }\mathbb{E}\left[ D\left(
t,T_{A}\right) \mathbf{1}_{\left\{ t<\hat{\tau}\leq T_{A}\right\} }\left(
1-R\right) |\mathcal{\hat{H}}_{t}\right]  \\
&&+\mathbf{1}_{\left\{ \hat{\tau}\leq t\right\} }\left( 1-R\right) P\left(
t,T_{A}\right)  \\
&=&\hat{O}_{1}+\hat{O}_{2}+\hat{O}_{3}
\end{eqnarray*}
We might have selected a different martingale dynamics, including any smile
dynamics, since our results are completely general. We stick to lognormality
for consistency with market standards in the pricing of credit options.

\subsection{Numeraire Pricing\label{Section_Numeraire}}

The $T_{A},T_{M}$-no-armageddon pricing measure $\mathbb{\hat{Q}}%
^{T_{A},T_{M}}$ introduced above is all we need for arbitrage-free pricing
of Index Options. However, it has been defined only on $\mathbb{\hat{Q}}%
^{T_{A},T_{M}}$ and only $\mathcal{\hat{H}}_{t}$-martingales have been
considered for developing a pricing formula. This is a strong restriction if
we want to extend the above results to more general products, that may
depend on the joint dynamics of the Index spread and other quantities, or on
the dynamics of the Index spread under different measures. Therefore it
appears convenient to introduce an extension of $\mathbb{\hat{Q}}%
^{T_{A},T_{M}}$ to the global $\sigma $-algebra $\mathcal{F}$, and it would
be particularly advantageous if this extension could be defined as a measure
associated to a standard \emph{numeraire}, as in Geman et al. (1996).

The $T_{A},T_{M}$-no-armageddon pricing measure $\mathbb{\hat{Q}}%
^{T_{A},T_{M}}$ introduced above has been defined and used in a way similar
to the numeraire measures introduced by Geman et al. (1995) and Jamshidian
(1997). However, it is not a numeraire measure, since the quantity $\hat{\Pi}%
\left( \gamma _{t}^{T_{A},T_{M}}\right) $ used to define this measure is not
the price of a tradable asset. In fact $\hat{\Pi}\left( \gamma
_{t}^{T_{A},T_{M}}\right) $ is an expectation with respect to $\mathcal{\hat{%
H}}_{t}$, a filtration that does not include all available market
information. This is why we went through such a detailed derivation.

For having a numeraire measure we need the Radon-Nikodym derivative to be
defined using prices of tradable assets, which by no-arbitrage need to be $%
\mathcal{F}_{t}$ expectation of $\mathcal{F}$-measurable claims. Such a
definition implies all the properties typical of the numeraire measures,
such as the fact that prices of tradable assets expressed in terms of a
numeraire $N$ are $\mathcal{F}_{t}$-martingales under the associated $%
\mathbb{Q}^{N}$ measure.

The issue of extending a measure initially given on a subfiltration of the
filtration $\left( \mathcal{F}_{t}\right) _{t\geq 0}$, and to define this
extension as a probability measure associated to a standard numeraire, is
dealt with for the single name case by Jamshidian (2004). With reference to
Jamshidian (2004) for details on the analytical derivations, in this section
we briefly see how this extension can be performed for $\mathbb{\hat{Q}}%
^{T_{A},T_{M}}$, adapting results to our context (therefore with $\hat{\tau}$
stopping time replacing the single name stopping time $\tau $).

The quantity
\begin{equation*}
\gamma ^{T_{A},T_{M}}=\sum_{j=A+1}^{M}\frac{B}{B_{T_{j}}}\alpha _{j}\left( 1-%
\frac{L\left( T_{j}\right) }{\left( 1-R\right) }\right) ,
\end{equation*}
where $B=B_{\bar{T}}$, is called \emph{prenumeraire}. Notice it is a
terminal payoff, corresponding to the index annuity observed at the terminal
date in our time horizon. Notice that $\hat{\Pi}\left( \gamma
_{T_{A}}^{T_{A},T_{M}}\right) $, the quantity we used for defining $\mathbb{%
\hat{Q}}^{T_{A},T_{M}}$, is
\begin{equation*}
\hat{\Pi}\left( \gamma _{T_{A}}^{T_{A},T_{M}}\right) =B_{T_{A}}\mathbb{E}%
\left[ \frac{\gamma ^{T_{A},T_{M}}}{B}|\mathcal{\hat{H}}_{T_{A}}\right] .
\end{equation*}
Jamshidian (2004) now suggests to choose as a \emph{numeraire} any claim $%
C^{T_{A},T_{M}}$ such that its $T_{A}$ fair price is
\begin{equation}
C_{T_{A}}^{T_{A},T_{M}}=B_{T_{A}}\mathbb{E}\left[ \frac{C^{T_{A},T_{M}}}{B}|%
\mathcal{F}_{T_{A}}\right] =\hat{\Pi}\left( \gamma
_{T_{A}}^{T_{A},T_{M}}\right) ,
\label{formula_numeraireprenumeraire_consistency_condition}
\end{equation}
and to define on $\mathcal{F}$ a measure $\mathbb{\hat{Q}}^{T_{A},T_{M}}$ as
\begin{equation}
\frac{d\mathbb{\hat{Q}}^{T_{A},T_{M}}}{d\mathbb{Q}}=\frac{%
B_{0}\;C^{T_{A},T_{M}}}{C_{0}^{T_{A},T_{M}}\;B}.
\label{formula_randonnikodym_NOrestriction}
\end{equation}
This is sufficient to define an associated probability measure such that its
restriction to $\mathcal{\hat{H}}_{T_{A}}$ does not depend on the choice of $%
C^{T_{A},T_{M}}$ as long as (\ref
{formula_numeraireprenumeraire_consistency_condition}) is satisfied. This
restriction, in particular, coincides with $\mathbb{\hat{Q}}^{T_{A},T_{M}}$
as defined on $\mathcal{\hat{H}}_{T_{A}}$ in Section \ref
{Section_equivalentmeasure}, thus justifying the notation used (the easy
proof is left to the reader). A very natural choice that we propose is to
choose
\begin{equation*}
C^{T_{A},T_{M}}=\frac{B}{B_{T_{A}}}\hat{\Pi}\left( \gamma
_{T_{A}}^{T_{A},T_{M}}\right) \text{.}
\end{equation*}
With this choice, the Radon-Nikodym derivative of $\mathbb{\hat{Q}}%
^{T_{A},T_{M}}$ with respect to the risk neutral measure is
\begin{equation*}
\frac{d\mathbb{\hat{Q}}^{T_{A},T_{M}}}{d\mathbb{Q}}=\frac{B_{0}\;\hat{\Pi}%
\left( \gamma _{T_{A}}^{T_{A},T_{M}}\right) }{\hat{\Pi}\left( \gamma
_{0}^{T_{A},T_{M}}\right) \;B_{T_{A}}}
\end{equation*}
and trivially coincides with the Radon-Nikodym derivative computed in
Section \ref{Section_equivalentmeasure}.

Following Theorem 3.8 in Jamshidian (2004), always with $\hat{\tau}$
stopping time replacing the single name stopping time $\tau $, one can check
that, with this definition of the pricing measure, the pricing formula for
an Index option coincides with (\ref
{formula_optionpricing_beforelognormality}), with the difference that now
the pricing measure is $\mathbb{\hat{Q}}^{T_{A},T_{M}}$ defined by (\ref
{formula_randonnikodym_NOrestriction}), rather than only the restriction
defined by $Z_{T_{A}}=\left. \frac{d\mathbb{\hat{Q}}^{T_{A},T_{M}}}{d\mathbb{%
Q}}\right| _{\mathcal{\hat{H}}_{T_{A}}}$. The fundamental assumptions used
by Jamshidian (2004) for proving this result are that the model is \emph{%
complementary}, namely
\begin{equation*}
\mathcal{F}_{t}=\mathcal{\hat{J}}_{t}\vee \mathcal{\hat{H}}_{t},
\end{equation*}
and \emph{positive}, namely
\begin{equation*}
P^{\beta }\left( \hat{\tau}>t|\mathcal{\hat{H}}\right) >0\;a.s.
\end{equation*}
for any numeraire $\beta $. Both properties are natural in our framework.
Complementarity has already been explicitly assumed in the model definition,
while positivity had been assumed only for the risk neutral measure, but can
be with no harm extended to all numeraire measures.

Therefore in this setting the pricing formula (\ref
{formula_optionpricing_beforelognormality}) can be expressed through a
global, numeraire-based measure.

\section{Empirical Analysis\label{Section_Tests}}

In the previous part of this work we have provided a rigorous theoretical
framework for the pricing of Index options, that was missing in the market
approach. This has led to replacing the market option formula (\ref
{formula_wrong2_option_price}) with the arbitrage-free option formula (\ref
{formula_right_option_price}). In this section we point out that the
improvement of the arbitrage-free option formula (\ref
{formula_right_option_price}) on the market option formula (\ref
{formula_wrong2_option_price}) is not limited to the fact that (\ref
{formula_right_option_price}) has a rigorous justification that lacked in
the market approach.

Indeed, for options on particular portfolios, or for peculiar market
situations, the correct accounting of the portfolio armageddon risk avoids
painful mistakes in valuation. As a consequence of the crisis that struck
credit markets in summer 2007, related to the burst of the \emph{subprime}
loans-linked structured finance bubble, the market is currently in one of
those situations when the correct accounting of the portfolio armageddon
risk is crucial. In the following empirical tests we consider market data of
March 2007, before the subprime crisis, market data of August 2007, in the
middle of the crisis, and finally more recent data of December 2007. We
focus on options on the i-Traxx Europe Crossover index, that have a short
maturity (less than one year) and have been particularly traded and liquid
in recent years. Following the JPM quotation terminology, we name Call
Options the options on the Receiver side (that allow an investor to be \emph{%
long} on credit risk), and Put Options the options on the Payer side (that
allow an investor to be \emph{short} on credit risk).

\subsection{The data}

In Table 1 and in Figures 1 and 2 one sees market quotations for the main
reference credit market products. Notice that they include information on
Index Tranches, quoted through the so-called \emph{base correlation}. In
fact the arbitrage-free option formula (\ref{formula_right_option_price})
introduces an explicit dependence on default correlation information, that
instead does not enter directly the market option formula (\ref
{formula_wrong2_option_price}).

This comes from the definition of the spread in (\ref
{formula_right_spreaddefinition}) and from the $\hat{O}_{2}$ component of
the option price, that, under the assumption of independence of interest
rate and default risk, is
\begin{equation*}
\mathbf{1}_{\left\{ \hat{\tau}>t\right\} }\left( 1-R\right) P\left(
t,T_{A}\right) \mathbb{Q}\left( \hat{\tau}\leq T_{A}|\hat{\tau}>t,\mathcal{%
\hat{H}}_{t}\right) .
\end{equation*}
It depends on the conditional $\mathcal{\hat{H}}_{t}$-probability of a
portfolio armageddon in $\left( t,T_{A}\right] $ given that there was no
armageddon before $t$. Ceteris paribus, an armageddon event will be more
likely in a context of higher correlation. We have included also correlation
information from the CDX (American) market, since the two credit markets
have been historically very close.

\begin{figure}[h]
\includegraphics
[width=6cm,height=12cm,angle=-90]{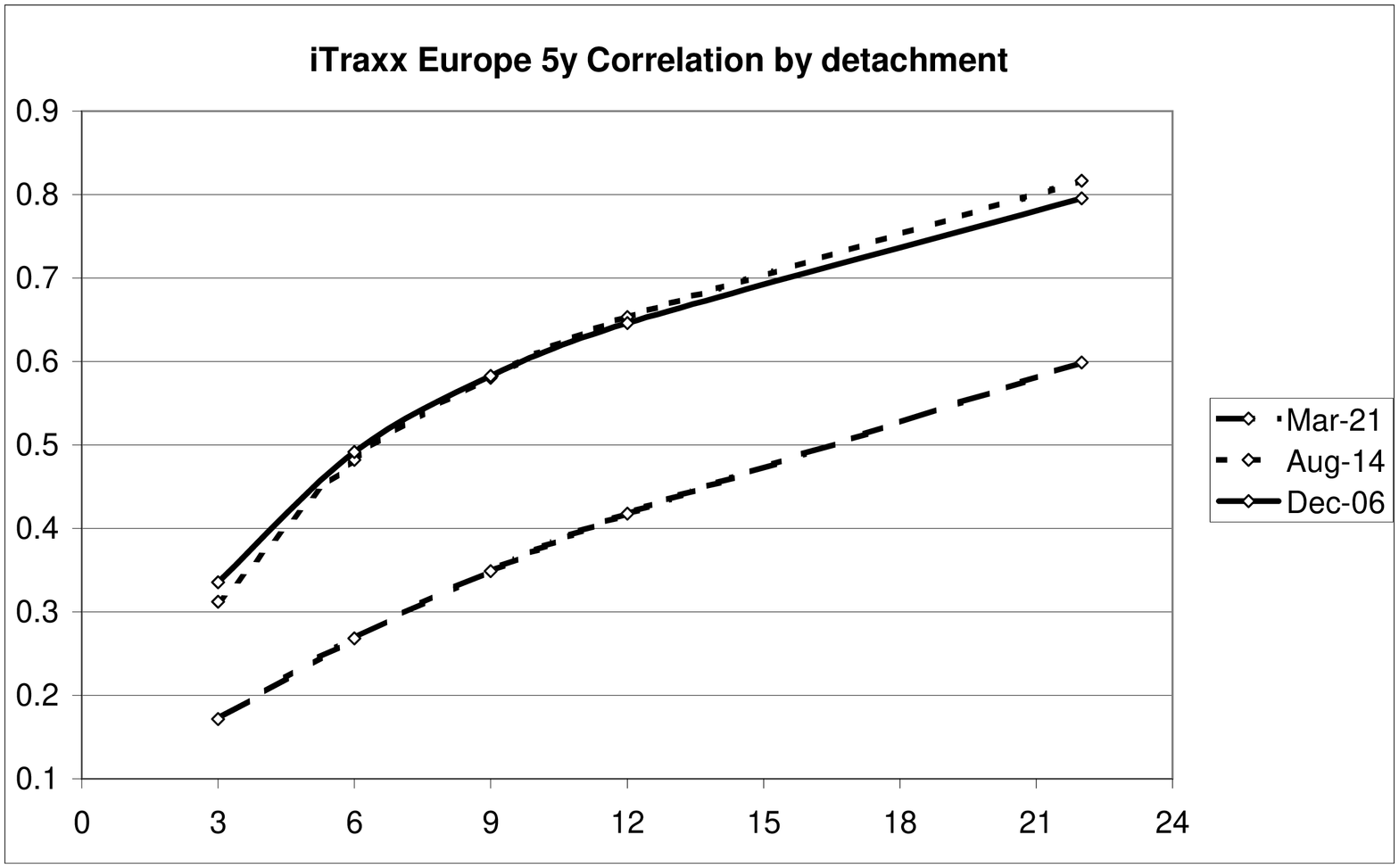}%
\hspace{1cm}
\caption{Quoted Base Correlation for Tranches on the iTraxx Europe Main
Index.}
\label{figure_base_corr_march}
\end{figure}
\begin{figure}[h]
\includegraphics
[width=6cm,height=11cm,angle=-90]{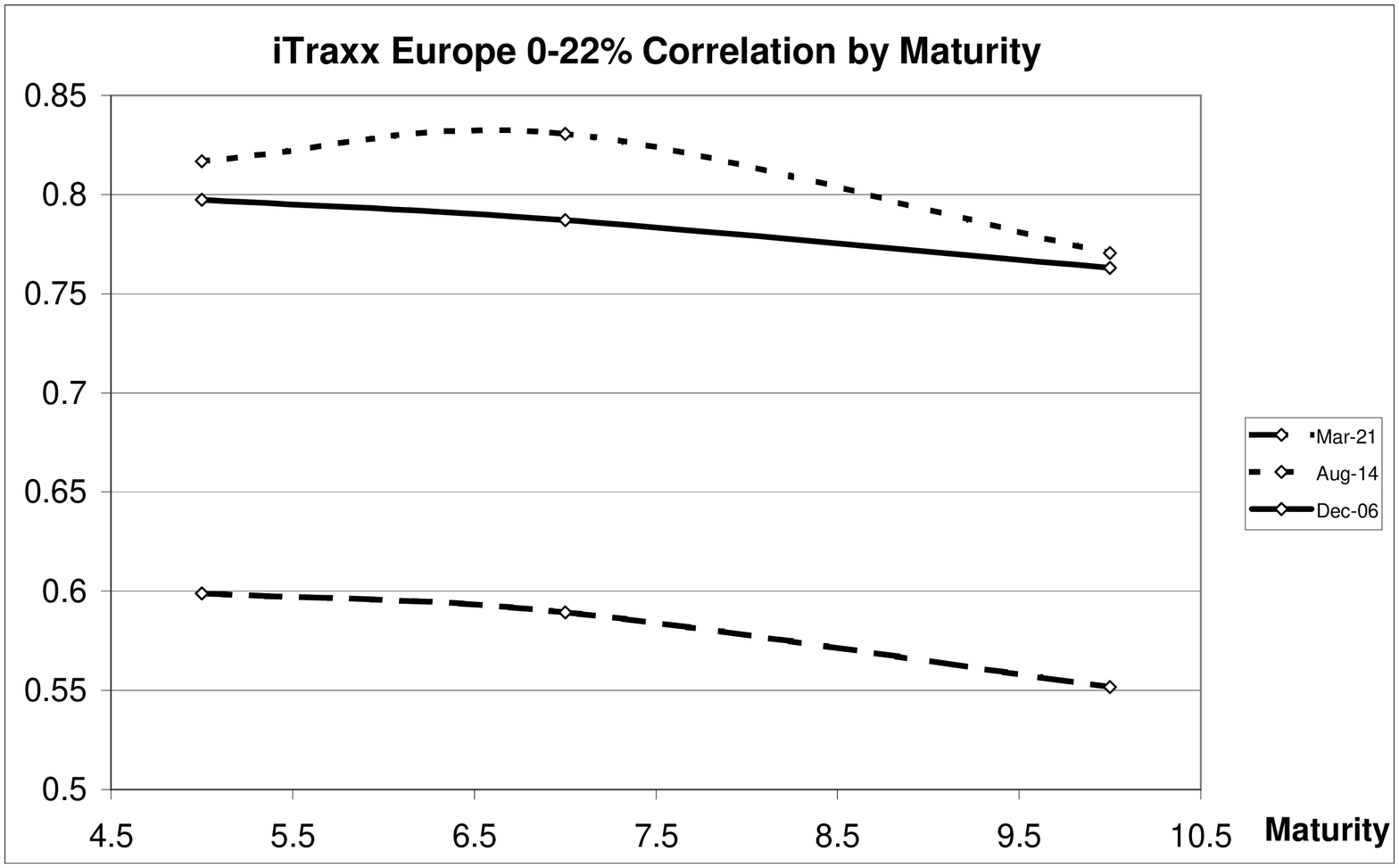}%
\hspace{1cm}
\caption{Quoted Base Correlation for Tranches on the iTraxx Europe Main
Index.}
\end{figure}

\noindent
\begin{tabular}{l}
\begin{tabular}{|l|l|l|l|l|}
\hline
& i-Traxx Main Spread (bps) & i-Traxx Crossover Spread (bps) & i-Traxx $\rho
_{22\%}$ & CDX $\rho _{30\%}$ \\ \hline
March 21, 2007 & 25.5 & 219 & 0.60 & 0.74 \\ \hline
August 14, 2007 & 53 & 361 & 0.82 & 0.95 \\ \hline
December 6, 2007 & 56 & 350 & 0.80 & 0.97 \\ \hline
\end{tabular}
\bigskip \\
Table 1: Credit market data (5y mat) on the 3 trading days considered.
\end{tabular}
\bigskip

A brief explanation of the correlation quotation system, and of how it is
used in market practice is here in order. The credit market associates to
each portfolio tranche $\left[ 0,x\right) $, with \emph{detachment} $x$
expressed in percentage points, a correlation value $\rho _{x}$, called
\emph{base correlation}. This value, together with the default intensities
of the single names, allows to compute the price of a tranche with the
market standard Gaussian Copula conventional model, for which we refer for
example to Glasserman et al. (2007) and references therein. The fact that a
single correlation is provided is a consequence of the \emph{homogeneous
portfolio assumption}, implying that all names in a portfolio have identical
features, including equal interdependence across all couples of issuers in
the pool. This provides a relatively easy and intuitive valuation and
quotation method, that we comply with in the following. We may use some
advanced loss model such as the GPL model of Brigo et al. (2006), able to
calibrate consistently market tranches. We prefer to follow the Gaussian
Copula since it is a recognized market standard.

For pricing i-Traxx Crossover Index Options we need computing $\hat{\tau}$
probabilities, thus one needs the correlations associated to the most senior
tranche possible: the one that is affected when, and only when, all
portfolio issuers default. It is the tranche that covers losses from a
fraction $\frac{\left( n-1\right) \left( 1-R\right) }{n}$ to a fraction $%
\left( 1-R\right) $ of the portfolio notional. For the 50 name Crossover
Index, with 40\% recovery, these detachment points correspond to $\left[
0,x\right) $ tranches where $x\in \left\{ 58.8\%;60\%\right\} $. We have two
problems in terms of market availability of the data. First, market
quotations are given only up to $x=22\%$. Second, correlation is not quoted
directly for the i-Traxx Crossover but only for the i-Traxx Main Index. When
the correlation for a non-standard portfolio, or for a non-standard
detachment, is not provided by the market, it is common practice (see Baheti
and Morgan (2007)) to use interpolation/extrapolation and \emph{mapping}
techniques for obtaining these correlations from the array of available base
correlations in Figure 1. However, due to lack of standardization and
theoretical justification, we prefer to take a more conservative and robust
approach, explained in the following.

It is market common agreement that correlation clearly increases with size
of equity tranches, as one can see in Figure 1. In the i-Traxx market there
is also a less marked tendency of correlation to decrease with maturity (see
Figure 2). We are interested in tranches $\left\{ 58.8\%;60\%\right\} $,
with maturities lower than one year. Therefore, according to both market
tendencies visible on iTraxx data, and in particular to the first one, we
should consider a correlation higher than $\rho _{22\%}$ of Table 1, the
i-Traxx correlation for the largest tranche and shortest maturity quoted. We
may expect a correlation even higher than the highest level quoted by the
CDX market, $\rho _{30\%}$.

However we do not increase correlation further, but instead we consider a
range of equally spaced correlations in-between the most senior base
correlations of i-Traxx ($\rho _{22\%}$) and CDX ($\rho _{30\%}$). Thus,
while extrapolation would suggest, for the seniority we need, to increase
correlation beyond any quoted $\rho $, we limit this correlation to the
market quoted CDX $\rho _{30\%}$. This has two advantages. First we are
robust towards a possible limit to the increase of correlation, due to the
fact that the Crossover is a less senior Index.\footnote{%
As explained in Baheti and Morgan (2007), a less senior index such as the
Crossover Index is more subject to idiosyncratic rather than systemic risk,
thus correlation should be somewhat reduced. However, analysis of the most
senior index in the North America market, the CDX.NA.IG, and the most
junior, CDX.NA.HY, shows that this difference is very low for very senior
tranches such as those we are interested in. Indeed, correlation is high for
very senior tranches to represent that their risk is related mainly to
systemic credit events. Such events, by definition, do not depend on the
specific portfolio considered.} Secondly, our choice tends to \emph{%
underestimate} the probability of $\hat{\tau}$, compared to the standard
market approach of extrapolating correlations. This implies also that, in
the following tests, the difference between our arbitrage-free pricing of
Index Options and the standard Market pricing is more likely to be
underestimated, rather than overestimated.

\subsection{Results: before and after the Subprime crisis}

Consider now Tables 2 and 3, reporting our Credit Index option pricing
results. In the upper subtables we see results for the 20 December 2007
maturity index option on March 21 2007. In the first row is the price of a
standard Index Option using the market option formula (\ref
{formula_wrong2_option_price}). This price is fully consistent with market
quotations, that include both a price quotation and an implied volatility
quotation. From the implied volatility quotation one can compute the price
through the market option formula (\ref{formula_wrong2_option_price}), and
check that it coincides with the price quotation.

Below is the price computed with the arbitrage-free option formula (\ref
{formula_right_option_price}), using the same implied volatility provided by
market quotations, and with the different correlation values considered.

The two formulas give very similar results, with differences around 1 bp or
even lower. Such a difference does not appear to be relevant under a
financial point of view, also considering that, as shown in Table 4, for the
20 December 2007 Index Option the upside part of the bid/offer spread
(difference between \emph{ask price} and \emph{mid price}) for Put options
is always more than $10$ bps and the downside part of the bid/offer spread
(difference between mid price and \emph{bid price}) for the Call option is
smaller but in any case larger than $1$ bps. Notice that the upside part is
the relevant one when the No-Arbitrage option overprices the market quotes,
and viceversa. This appears to be a good estimate of the threshold over
which a pricing difference starts to be financially relevant.

Now we move to August 14, 2007. We see in Table 1 that, although the pitch
of the credit indices rally was already over (Crossover spread touched 463
bps at the end of July 2007), the market situation has changed radically
compared to March 2007. Both Index spreads and Tranche correlations have
dramatically increased.

Taking into account the new market conditions shown in Table 1, we evaluate
a 9 month maturity option, thus corresponding to the option evaluated on
March 21. Results are presented in the mid subtables of Tables 2 and 3. Here
the difference between the price computed with the market option formula (%
\ref{formula_wrong2_option_price}) and the price with the arbitrage-free
option formula (\ref{formula_right_option_price}) is much larger. Looking at
Call Options in Table 2, we see that the option with the lowest strike has a
price with the arbitrage-free option formula (\ref
{formula_right_option_price}) which is little more than a half of the price
computed with the market option formula (\ref{formula_wrong2_option_price}),
because the perceived higher systemic risk, triggered by the subprime
crisis, has made the risk-neutral probability of an armageddon event not
negligible anymore.

The price differences appear particularly large, under a financial point of
view, if compared to the values of Table 4, that reports the appropriate
market bid-ask quotations for the longest maturity Index Options quoted in
any of the three trading days considered. Indeed, for all strikes and all
possible correlation values, the difference between market pricing and no
arbitrage pricing is higher than the appropriate bid-ask spread.

\emph{Therefore taking correctly into account the possibility of portfolio
armageddon is not only an issue that allows the definition of the spread and
of the pricing measure to be regular under a mathematical point of view, but
for options on the Crossover Index it is also of financial relevance in some
peculiar market situations, such as the credit crunch of summer 2007.}%
\newpage

\begin{tabular}{l}
{\textbf{March 21, 2007}} \\
\begin{tabular}{|l|r|r|r|r|r|}
\hline
Strike & 200 & 225 & 250 & 275 & 300 \\ \hline
\textbf{Market Formula} & \textbf{73.72} & \textbf{115.32} & \textbf{165.57}
& \textbf{226.46} & \textbf{291.95} \\ \hline
No-Arbitrage Formula \textbf{\ }$\rho =0.6$ & 73.61 & 115.22 & 165.48 &
226.38 & 291.88 \\ \hline
No-Arbitrage Formula \textbf{\ }$\rho =0.65$ & 73.44 & 115.06 & 165.34 &
226.26 & 291.77 \\ \hline
No-Arbitrage Formula \textbf{\ }$\rho =0.7$ & 73.07 & 114.73 & 165.04 &
225.99 & 291.53 \\ \hline
No-Arbitrage Formula \textbf{\ }$\rho =0.75$ & 72.25 & 113.99 & 164.38 &
225.40 & 291.00 \\ \hline
\end{tabular}
\smallskip \\
{\textbf{August 14, 2007}} \\
\begin{tabular}{|l|r|r|r|r|r|}
\hline
Strike & 300 & 325 & 350 & 375 & 400 \\ \hline
\textbf{Market Formula} & \textbf{120.56} & \textbf{155.38} & \textbf{196.01}
& \textbf{240.53} & \textbf{290.56} \\ \hline
No-Arbitrage Formula \textbf{\ }$\rho =0.8$ & 113.14 & 148.27 & 189.20 &
234.00 & 284.30 \\ \hline
No-Arbitrage Formula \textbf{\ }$\rho =0.85$ & 106.09 & 141.51 & 182.72 &
227.80 & 278.36 \\ \hline
No-Arbitrage Formula \textbf{\ }$\rho =0.9$ & 92.33 & 128.32 & 170.10 &
215.72 & 266.79 \\ \hline
No-Arbitrage Formula \textbf{\ }$\rho =0.95$ & 63.06 & 100.31 & 143.33 &
190.13 & 242.29 \\ \hline
\end{tabular}
\smallskip \\
{\textbf{Dec 6, 2007}} \\
\begin{tabular}{|l|r|r|r|r|r|}
\hline
Strike & 325 & 350 & 375 & 400 & 425 \\ \hline
\textbf{Market Formula} & \textbf{123.38} & \textbf{162.10} & \textbf{205.62}
& \textbf{253.52} & \textbf{305.32} \\ \hline
No-Arbitrage Formula \textbf{\ }$\rho =0.79$ & 117.27 & 156.32 & 200.18 &
248.40 & 300.50 \\ \hline
No-Arbitrage Formula \textbf{\ }$\rho =0.85$ & 110.37 & 149.81 & 194.05 &
242.64 & 295.09 \\ \hline
No-Arbitrage Formula \textbf{\ }$\rho =0.91$ & 92.87 & 133.32 & 178.54 &
228.07 & 281.41 \\ \hline
No-Arbitrage Formula \textbf{\ }$\rho =0.97$ & 49.61 & 92.68 & 140.44 &
192.38 & 247.99 \\ \hline
\end{tabular}
\smallskip \smallskip \\
Table 2: Prices in bps of Call Options on i-Traxx Crossover 5y On-the-Run.
Maturity 9m
\end{tabular}

\bigskip

\bigskip
\begin{tabular}{l}
{\textbf{March 21, 2007}} \\
\begin{tabular}{|l|r|r|r|r|r|}
\hline
{Strike} & 200 & 225 & 250 & 275 & 300 \\ \hline
{\textbf{Market Formula}} & {\textbf{299.73}} & {\textbf{250.74}} & {\textbf{%
210.42}} & {\textbf{180.73}} & {\textbf{155.64}} \\ \hline
{No-Arbitrage Formula \textbf{\ }$\rho =0.6$} & 299.76 & 250.78 & 210.47 &
180.79 & 155.70 \\ \hline
{No-Arbitrage Formula \textbf{\ }$\rho =0.65$} & 299.80 & 250.85 & 210.54 &
180.88 & 155.81 \\ \hline
{No-Arbitrage Formula \textbf{\ }$\rho =0.7$} & 299.90 & 250.98 & 210.71 &
181.08 & 156.04 \\ \hline
{No-Arbitrage Formula \textbf{\ }$\rho =0.75$} & 300.13 & 251.28 & 211.09 &
181.53 & 156.55 \\ \hline
\end{tabular}
\smallskip \\
{\textbf{August 14, 2007}} \\
\begin{tabular}{|l|r|r|r|r|r|}
\hline
{Strike} & 300 & 325 & 350 & 375 & 400 \\ \hline
{\textbf{Market Formula}} & {\textbf{559.60}} & {\textbf{519.46}} & {\textbf{%
485.13}} & {\textbf{454.69}} & {\textbf{429.76}} \\ \hline
{No-Arbitrage Formula \textbf{\ }$\rho =0.8$} & 561.20 & 521.36 & 487.33 &
457.18 & 432.52 \\ \hline
{No-Arbitrage Formula \textbf{\ }$\rho =0.85$} & 562.75 & 523.20 & 489.46 &
459.58 & 435.18 \\ \hline
{No-Arbitrage Formula \textbf{\ }$\rho =0.9$} & 565.85 & 526.88 & 493.70 &
464.37 & 440.47 \\ \hline
{No-Arbitrage Formula \textbf{\ }$\rho =0.95$} & 572.82 & 535.11 & 503.17 &
475.02 & 452.22 \\ \hline
\end{tabular}
\smallskip \\
{\textbf{Dec 06, 2007}} \\
\begin{tabular}{|l|l|l|l|l|l|}
\hline
Strike & 325 & 350 & 375 & 400 & 425 \\ \hline
{\textbf{Market Formula}} & \textbf{452.15} & \textbf{410.36} & \textbf{%
373.37} & \textbf{340.75} & \textbf{312.04} \\ \hline
{No-Arbitrage Formula \textbf{\ }$\rho =0.79$} & 453.88 & 412.43 & 375.77 &
343.48 & 315.07 \\ \hline
{No-Arbitrage Formula \textbf{\ }$\rho =0.85$} & 455.86 & 414.80 & 378.53 &
346.60 & 318.53 \\ \hline
{No-Arbitrage Formula \textbf{\ }$\rho =0.92$} & 461.09 & 421.03 & 385.74 &
354.75 & 327.58 \\ \hline
{No-Arbitrage Formula \textbf{\ }$\rho =0.98$} & 475.32 & 437.89 & 405.14 &
376.56 & 351.66 \\ \hline
\end{tabular}
\smallskip \smallskip \\
Table 3: Prices in bps of Put Options on i-Traxx Crossover 5y On-the-Run.
Maturity 9m.
\end{tabular}

\begin{tabular}{l}
\begin{tabular}{|l|l}
\cline{1-1}
March 21 &
\begin{tabular}{|l|r|r|r|r|r|}
\hline
Strike & 200 & 225 & 250 & 275 & 300 \\ \hline
Mid-Bid for Call options & 8 & 6 & 4 & 2 & 2 \\ \hline
Ask-Mid for Put options & 20 & 16 & 14 & 14 & 15 \\ \hline
\end{tabular}
\\ \cline{1-1}
August 14 &
\begin{tabular}{|l|r|r|r|r|r|}
\hline
Strike & 300 & 325 & 350 & 375 & 400 \\ \hline
Mid-Bid for Call options & 7 & 5 & 3 & 1 & 1 \\ \hline
Ask-Mid for Put options & 11 & 12 & 13 & 17 & 20 \\ \hline
\end{tabular}
\\ \cline{1-1}
December 6 &
\begin{tabular}{|l|r|r|r|r|r|}
\hline
Strike & 325 & 350 & 375 & 400 & 425 \\ \hline
Mid-Bid for Call options & 4 & 1 & 0.1 & 0.1 & 0.3 \\ \hline
Ask-Mid for Put options & 8 & 7 & 6 & 7 & 9 \\ \hline
\end{tabular}
\\ \cline{1-1}
\end{tabular}
\smallskip \smallskip \\
Table 4. Market Bid-Ask on quoted Index Options.
\end{tabular}
\bigskip

For Put options of Table 3, price differences appear more moderate, since
only for the highest correlation considered (corresponding to the CDX
correlation\footnote{%
The correlation values we report for the CDX market on August and December
2007 may appear exceptionally high, however we point out that $\rho _{30\%}$
for CDX exceeded $99\%$ on the last trading week of November 2007.}) the
difference exceeds the relevant bid-ask, in spite of the fact that in some
cases such differences are higher than 20 bps.

One may reasonably argue that the liquidity of the market in August 2007 was
particularly low, and the situation, being the immediate aftermath of an
unexpected crisis, was extremely peculiar and not likely to last for any
relevant time or to appear again in the market.

In order to verify this argument, we wait until December 2007, and we price
a 9-month option taking into account the December 06, 2007, market
situation. The results appear to belie the above argument. In spite of the
slightly lower Crossover spreads, the market still discounts a high
perceived systemic risk, expressed by extremely high base correlation
quotes. As a consequence, the correct accounting of the market-implied
risk-neutral probability of a portfolio armageddon is still very relevant.
This is shown both by Call prices, where the differences range from 15\% to
almost 60\% of the Market Formula Price, and by Put quotations, where the
number of cases when the difference between the two formulas exceeds the
available bid-ask is even higher than in August 14. This relevant impact on
option prices of the risk of a systemic crisis appears one of the possible
reasons for the fact that the liquidity in the credit option market, after
summer 2007, is concentrated on the shortest maturities.

We also mention that, although the historical probability of total portfolio
default appears clearly negligible when we are considering large portfolios
of investment grade issuers (such as the i-Traxx 125-name Main Index), there
is evidence in the literature that also this risk is priced, so that its
risk-neutral probability is not negligible. An example is in Brigo et al.
(2006), where it is shown that the Generalized Poisson Loss Model needs to
include a jump process associated to an armageddon event, if one wants to
price correctly market tranches of different subordination and maturity
written on the iTraxx Main Index.

Additionally, there is a category of credit portfolio options for which,
even in normal market conditions, the probability of the portfolio to be
wiped out by defaults is never negligible, even in normal market conditions:
it is the case of Tranche options, in particular for the most risky equity
or mezzanine tranches. For such options, that are often embedded in
cancelable tranches, this issue is always crucial in all market situations.
Thus, for an option on a $\left[ 0,x\right) $ Tranche, using a formula
analogous to the arbitrage-free option formula (\ref
{formula_right_option_price}) is fundamental for correct pricing in all
market situations. Clearly, for a tranche option, the Tranched Loss
\begin{equation*}
L_{x,y}\left( t\right) =\frac{1}{y-x}\left[ \left( L\left( t\right)
-x\right) ^{+}-\left( L\left( t\right) -y\right) ^{+}\right]
\end{equation*}
must replace the Loss $L\left( t\right) $ and $\hat{\tau}$ is the first time
when $L_{x,y}\left( t\right) =1$.

\section{Conclusion}

In this work we have considered three fundamental problems of the standard
market approach to the pricing of portfolio credit options: the definition
of the index spread is not valid in general, the payoff considered leads to
a pricing which is not defined in all states of the world, the candidate
numeraire to define a pricing measure is not strictly positive, which would
lead to an non-equivalent pricing measure.

We have given to the three problems a general mathematical solution, based
on a novel way of modelling the flow of information through the definition
of a new subfiltration. Using this subfiltration, we take into account
consistently the possibility of default of all names in the portfolio, that
is neglected in the standard market approach. We have shown that, while this
mispricing can be negligible for standard options in normal market
conditions, it can become highly relevant for less standard options or in
stressed market conditions. In particular, we show on 2007 market data that
after the subprime credit crisis the mispricing of the market formula
compared to the no arbitrage formula we propose has become financially
relevant even for the liquid Crossover Index Options.

The definition of an equivalent pricing measure through subfiltrations, and
the related results on the dynamics of a well defined portfolio credit
spread, lay the foundations of an extension to a multiname credit setting of
the no-arbitrage approach known as Market Models, dating back to Brace,
Gatarek and Musiela (1997) and Jamshidian (1997). Such extension is
considered by Jamshidian (2004), Brigo (2005) and Brigo and Morini (2005)
for the case of single name credit products. Here, instead, we consider the
extension to a multiname setting, that is technically from the case of
single name modelling, and also shows a different market applicability. In
fact the Market Models approach is characterized by the fact of allowing
precise arbitrage-free consistency in the modelling of market rates and
spreads, and of requiring information on volatilities and correlations of
such rates and spreads. These features have made the approach very
successful in the interest rate derivatives market, but they hardly hold for
single name credit derivatives, split among different reference with rare
options. Instead, the reference iTraxx and CDX Indices now absorb the larger
part of the credit derivative market, providing a reference market from
which information for modelling market quantities can be extracted. Even
after the credit spread rally in summer 2007, multiname indices experienced
a return to trading activity while the single name credit market was still
dried out.

\end{document}